\documentclass[preprint2]{aastex}

\usepackage{amsmath}
\usepackage{rotating}

\newcommand{\ewHb}{$EW_{{\rm H}_\beta}$}
\newcommand{\fe}{$\overline{EW}_{\rm Fe}$}
\newcommand{\teff}{$T_{\rm{eff}}$}
\newcommand{\logg}{log$g$}
\newcommand{\feh}{[Fe/H]}
\newcommand{\fehlm}{[Fe/H]$_{LM}$}
\newcommand{\fehlmd}{[Fe/H]$_{LM2D}$}

\shorttitle{The K giant stars from the LAMOST survey I}
\shortauthors{Liu et al.}

\begin{document}


\title{The K giant stars from the LAMOST survey data I: identification, metallicity, and distance}


\author{Chao Liu\altaffilmark{1}, Li-Cai Deng\altaffilmark{1}, Jeffrey L. Carlin\altaffilmark{2}, Martin C. Smith\altaffilmark{3}, Jing Li\altaffilmark{1,3}, Heidi Jo Newberg\altaffilmark{2}, Shuang Gao\altaffilmark{1}, Fan Yang\altaffilmark{1}, Xiang-Xiang Xue\altaffilmark{4}, Yan Xu\altaffilmark{1}, Yue-Yang Zhang\altaffilmark{1}, Yu Xin\altaffilmark{1}, Ge Jin\altaffilmark{5}}\email{liuchao@nao.cas.cn}


\altaffiltext{1}{Key Laboratory of Optical Astronomy, National Astronomical Observatories, Chinese Academy of Sciences, Datun Road 20A, Beijing 100012, China}
\altaffiltext{2}{Department of Physics, Applied Physics and Astronomy, Rensselaer Polytechnic Institute, 110 8th Street, Troy, NY 12180, USA}
\altaffiltext{3}{Shanghai Astronomical Observatory, Chinese Academy of Sciences, 80 Nandan Road, Shanghai 200030, China}
\altaffiltext{4}{Max Planck Institute for Astronomy, K\"onigstuhl 17, Heidelberg D-69117, Germany}
\altaffiltext{5}{University of Science and Technology of China, Hefei 230026, China}


\begin{abstract}
We present a support vector machine classifier to identify the K giant stars from the LAMOST survey directly using their spectral line features. The completeness of the identification is about 75\% for tests based on LAMOST stellar parameters. The contamination in the identified K giant sample is lower than 2.5\%.
Applying the classification method to about 2 million LAMOST spectra observed during the pilot survey and the first year survey, we select 298,036 K giant candidates. The metallicities of the sample are also estimated with uncertainty of $0.13\sim0.29$\,dex based on the equivalent widths of Mg$_{\rm b}$ and iron lines. 
A Bayesian method is then developed to estimate the posterior probability of the distance for the K giant stars, based on the estimated metallicity and 2MASS photometry. The synthetic isochrone-based distance estimates have been calibrated using 7 globular clusters with a wide range of metallicities. The uncertainty of the estimated distance modulus at $K=11$\,mag, which is the median brightness of the K giant sample, is about 0.6\,mag, corresponding to $\sim30$\% in distance.
As a scientific verification case, the trailing arm of the Sagittarius stream is clearly identified with the selected K giant sample. Moreover, at about 80\,kpc from the Sun, we use our K giant stars to confirm a detection of stream members near the apo-center of the trailing tail. These rediscoveries of the features of the Sagittarius stream illustrate the potential of the LAMOST survey for detecting substructures in the halo of the Milky Way. 
\end{abstract}



\section{Introduction}
The LAMOST {(the Large sky Area Multi-Object fiber Spectroscopic Telescope; also known as Guo Shou Jing Telescope)} project has carried out a pilot survey between October 2011 and June 2012 and obtained more than 700,000 spectra \citep{cui12,zhao12,lxw13}. The regular survey has operated since 2012 September and has already obtained about 2 million spectra as of June 2013. These spectra has been released as the DR1 catalog. A large fraction of them are K giant stars, which is of great interest in the studies of the Milky Way and in particular for the Galactic halo.

K giant stars are luminous and thus allow us to probe the Galaxy far beyond the solar neighborhood. Typically,  the absolute magnitude of K giant stars is between $M_r=2$ and -2\,mag. Given the limiting magnitude of $r=17.8$\,mag, the maximum distance at which LAMOST can detect K giant stars is $\sim$90\,kpc from the Sun.

\citet{xue12} found distances to more than 4000 K giant stars in SDSS/SEGUE survey with accuracy of $\sim$12\%. Because the limiting magnitude of SEGUE spectra is $g=20.2$\,mag, the maximum distance is much larger than that of LAMOST.
Although the K giant stars observed by LAMOST cannot compete with SDSS/SEGUE in terms of distance probed, the total number of this type of spectra in LAMOST will be two orders of magnitude larger than those in SDSS due to the huge {number of spectra} observed in the LAMOST survey and the fact that a larger fraction of bright stars are giants.

With such a huge dataset, we should be able to address many interesting and important questions on our galaxy, e.g., the total mass of the Milky Way, the shape of the dark matter halo, the kinematic substructures in the stellar spheroid, the mass distribution of the Galactic disk, the chemo-dynamical features and the evolution history of the disk, etc. Specifically, it will significantly improve the observational evidence of the kinematic substructures in the stellar halo \citep{deng12}. In the 2MASS and SDSS surveys, many substructures have been discovered in the past decades, including the Sagittarius dwarf galaxy stream \citep{ibata01,newberg02,majewski03}, Monoceros ring \citep{newberg02,yanny03}, Orphan stream \citep{belokurov06}, Virgo overdensity \citep{newberg02,vivas06,newberg07}, Triangulum-Andromeda overdensity \citep{majewski04b,rochapinto04}, Hercules-Aquila cloud \citep{belokurov07}, Cetus polar stellar stream \citep{newberg09}, Pisces stellar stream \citep{bonaca12,martin13}, and many other cold and weak streams \citep[e.g., ][]{grillmair06}. Although some of these substructures are prominent in photometric catalogs, a tiny fraction of their member stars have spectroscopic observations. The identification of the K giant members in known tidal streams will be crucial to constrain the orbits of the tidal streams and to constrain the merging history of their progenitors \citep[e.g.,][]{law05,law10}. Moreover, it can also be used to measure the total mass of the Milky Way \citep[e.g.,][]{koposov10}. 

In addition, the K giant stars can be used to discover new substructures which are otherwise not possible by any previous approaches. According to the $\Lambda$CDM cosmology, the halo of a Milky Way-like galaxy should contain hundreds of subhaloes, in which dwarf galaxies may be embedded. Current observations only find a few tidal streams and dozens of satellite dwarf galaxies around the Galaxy. This is the so called \emph{missing satellite problem}, which challenges all current theories \citep{klypon99,koposov08}. Because some of the merging dwarf galaxies form tidal streams during accretion, the search for new tidal substructure is one important way to address this discrepancy. Ultimately, the study of tidal streams will allow a better understanding on the formation history and evolution of a galaxy. 

In order to make optimal use of the K giant sample, a clean and relatively complete K giant catalog with distance estimates is highly desirable. In principle, K giant stars can be identified from measurements of stellar parameters, e.g. effective temperature (\teff) and surface gravity (\logg). However, estimations of the stellar parameters can only be reliably achieved for the \emph{good} spectra, i.e., the high signal-to-noise ratio data or the well flux {calibrated data}. As a result, a majority of K giant spectra with moderate or low signal-to-noise ratio, which are usually located at further distances, are missing from the parameter table. In order to reach {a} larger detection volume and hence maximize the scientific value of the LAMOST survey data, we take an alternative approach of identifying K giant stars directly from spectra instead of stellar parameters.
In addition to the identification of the K giant stars, the metallicity is necessary for the distance estimation. Consequently, one of the aims of the current work is to develop a more robust and reliable metallicity estimation method, especially for those spectra with low or moderate S/N. Finally, the distance of the K giant samples is determined with a robust statistical method.
 
A brief introduction to the data used in this work is given in section~\ref{sect:data}.
In section~\ref{sect:class}, a machine learning algorithm is developed for the identification of K giant stars from the stellar spectra of the LAMOST survey. The success of the classification method is verified by using MILES and SDSS public data. The method is then applied to the LAMOST data and a complete K giant dataset is produced.
In section~\ref{sect:metal}, a thin-plate spline model (hereafter, LM2D) for \feh\ estimation based on the Mg$_b$ and iron lines is introduced and applied to the identified K giant stars.
Subsequently, the distance of the selected K giant stars with low extinction is estimated based on the isochrone comparison in section~\ref{sect:dist}. 
The well-known kinematic substructure, the Sagittarius stream, is then identified from the K giant stars in section~\ref{sect:sgr}. Finally, a short conclusion of the current work is given in the last section. 

\section{LAMOST survey data}\label{sect:data}

The LAMOST DR1 catalog contains more than 2 million spectra, including about 700,000 spectra observed during the pilot survey. In total, there are about 1.9 million stellar spectra available from the LAMOST survey.

\begin{figure*}[htbp]
\begin{center}
\includegraphics[scale=0.75]{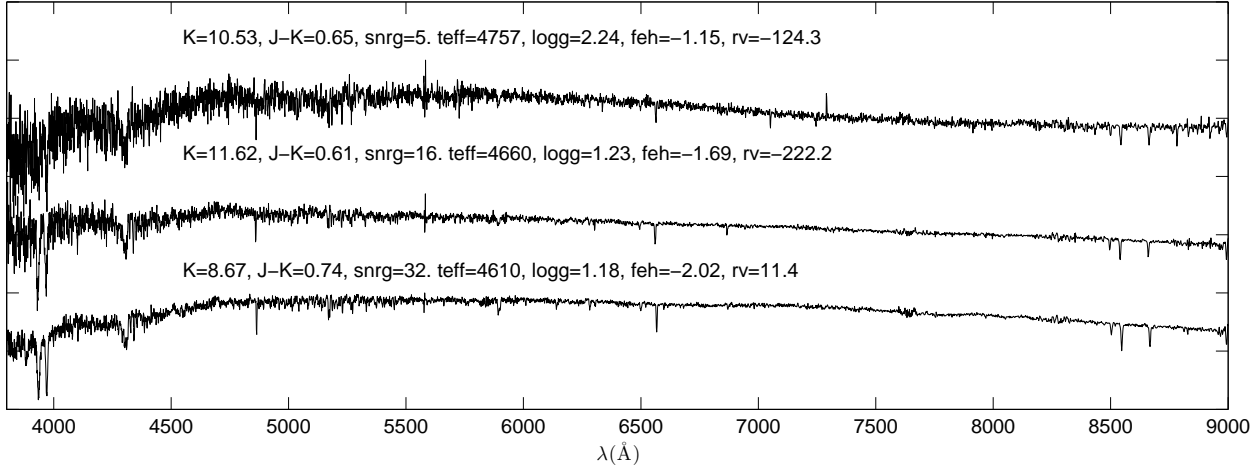}
\caption{Sample K giant spectra selected from LAMOST survey. The parameters, $K$ magnitude, $J-K$, signal-to-noise ratio in $g$ band (snrg), effective temperature (teff), surface gravity (logg), metallicity (feh), and radial velocity (rv), are marked above the corresponding spectrum.}\label{fig:samplespec}
\end{center}
\end{figure*}

The standard LAMOST pipeline \citep[][Luo et al. in preparation]{luo12} converts 2D into 1D spectra, corrects the flat field, combines the blue and red parts of the spectra, calibrates the wavelength and subtracts the sky background. Figure~\ref{fig:samplespec} {shows} sample spectra for K giant stars processed by the standard pipeline. 

For all spectra, the standard pipeline also provides the radial velocity based on the cross correlation {with} ELODIE library \citep{prugniel07}. However, only about half of the total data are high quality F, G, and K type spectra, for which stellar parameters are estimated using {\emph{Ulyss}} \citep{wu11} in the pipeline. {\emph{Ulyss} is a forward model method, which models the spectrum pixels as a linear combination of a set of non-linear components. Each non-linear component is a function of stellar parameters as well as radial velocity and it is defined in advance based on the ELODIE stellar library. The best fit stellar parameters and radial velocity of an observed stellar spectrum is determined iteratively by minimizing the $\chi^2$ value between the observed spectrum and the model.} {Because the stellar parameters of only about 50\% spectra with higher signal-to-noise ratio have been estimated, the selection function of the stars may be distorted and the sampling power of the survey may be weakened}

\section{K giant selection}\label{sect:class}

\subsection{Support vector machine classifier}
Support vector machine (SVM) is a machine learning algorithm which is suited for classification \citep{cortes95} and broadly used in {astronomy }\citep[e.g., ][]{bailerjones08,liu12,saglia12,bailerjones13}. As a supervised algorithm, it needs a set of known data to train the SVM model first. A subset of the training data are selected as the support vectors during the training phase. The support vectors define the linear boundary of classes in a high dimensional inner product space. When the training process is done, the SVM model is ready for prediction; any data input to the model will be marked as a certain class depending on the region to which the input data are projected. 

\subsection{Data preprocessing}
The full spectrum of a star does not carry useful information in every pixel. Some parts of the spectrum that contain strong line features play more important role in classification and parametrization. Therefore, we use the equivalent width ($EW$) of such strong lines to identify the K giant stars. 

\begin{figure*}[ht]
\begin{center}
\includegraphics[scale=0.7]{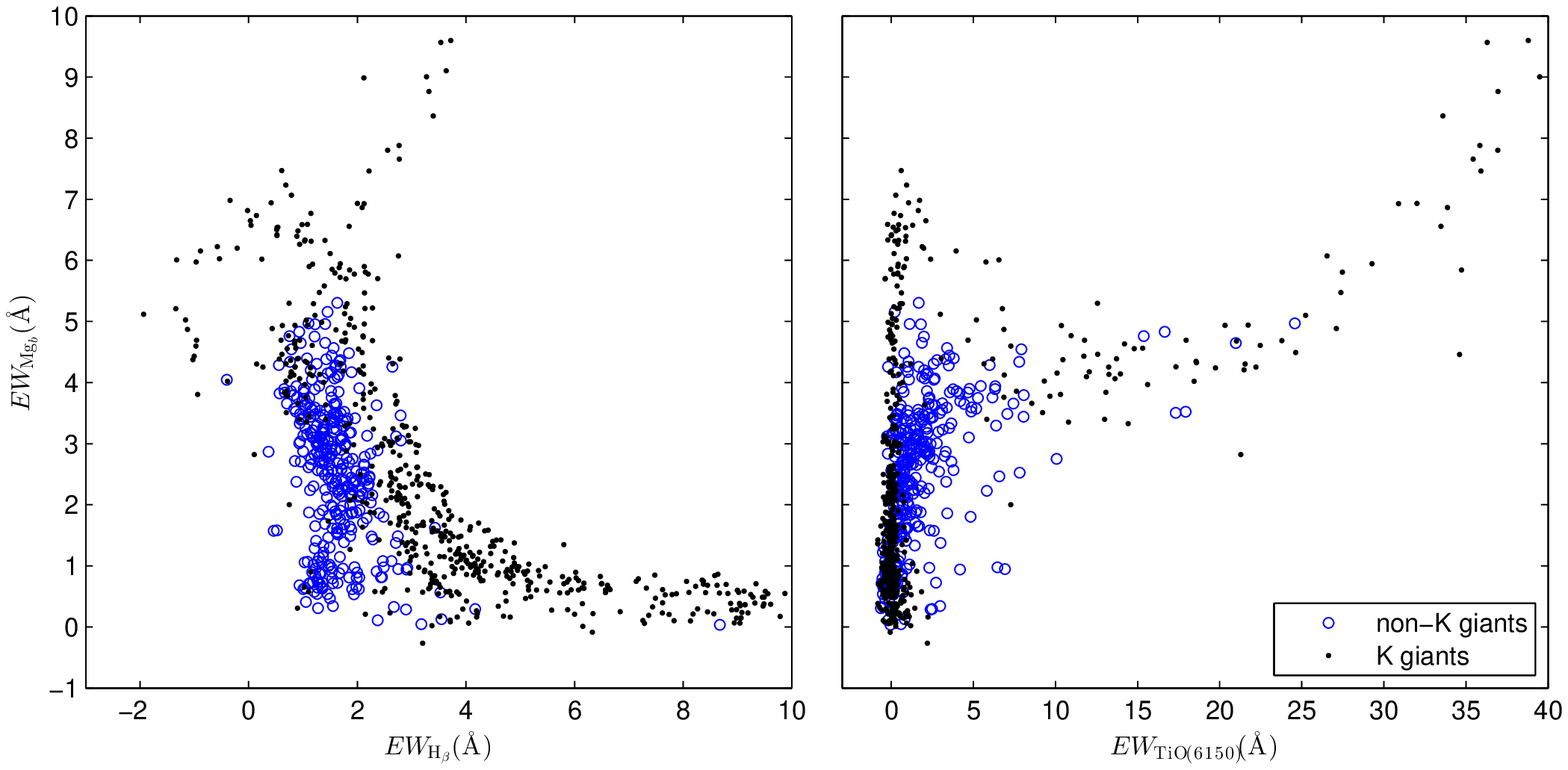}
\caption{The figure shows the $EW$s of Mg$_b$, H$_\beta$, and TiO (6150$\rm\AA$) for the MILES K giant stars (blue circles) and non-K giants (black dots). The left panel is $EW_{{\rm H}_\beta}$ vs. $EW_{{\rm Mg}_b}$ and the right panel is $EW_{{\rm TiO}(6150)}$ vs. $EW_{{\rm Mg}_b}$.}\label{fig:ewclass}
\end{center}
\end{figure*}

First, we select some of the Balmer lines, H$_\alpha$, H$_\beta$, H$_\gamma$, and H$_\delta$, as the indicators of temperature. However, for stars with \teff\ lower than $\sim5000$\,K, the Balmer lines are very weak or even invisible, so the TiO {feature near 6150$\rm\AA$} is used as temperature indicator in this case. Second, in order to distinguish the giant from dwarf stars, we need to use the lines that are sensitive to surface gravity. The magnesium lines around 5180$\rm\AA$ (including Mg$_1$, Mg$_2$, and Mg$_{\rm b}$) are good tracers for this purpose. Some other important features, e.g. CN, G band,  are also included. Finally, we use the $EW$s of the ten selected lines to characterize the whole spectrum. We adopt the Lick indexes\citep{worthey94} to measure the $EW$s using the following equation:
{
\begin{equation}
EW=\int{(1-{f_{line}(\lambda)\over{f_{cont}(\lambda)}})}{d\lambda},
\end{equation}
where $f_{cont}(\lambda)$ and $f_{line}(\lambda)$ are the fluxes of the continuum and the spectral line, respectively, both of which are functions of the wavelength $\lambda$.}

Figure~\ref{fig:ewclass} shows the difference between the K giant and non-K giant stars of the MILES library {\citep{miles}} in the $EW$s of Mg$_b$, H$_\beta$, and TiO. We can essentially separate the K giant stars from $EW_{{\rm H}_\beta}$ vs. $EW_{{\rm Mg}_b}$ with some local overlapping. TiO can help to distinguish the giant from the dwarf stars in some of the overlapped region, particularly at $4<EW_{{\rm Mg}_b}<5$.

\subsection{Training of the SVM classifier }\label{sect:trainsvm}
By training with a set of data with known giant/non-giant separation, the parameters within SVM can be properly tuned to obtain the best model for the classification.
For this purpose, the training data is defined using a common sample of the LAMOST pilot survey and SDSS DR9 \citep{ahn12} with the following additional criteria: i) the signal-to-noise ratio for SDSS spectra are larger than 20 and \teff, \logg, and \feh\ is provided by SSPP pipeline; ii) the spectra is marked as STAR in the LAMOST pipeline; and iii) the signal-to-noise ratio in $g$ band (denoted as S/N($g$)) measured from LAMOST spectra is larger than 10\footnote{The SDSS $g$ band filter covers from 4000 to 5300\,$\rm\AA$, containing most of the useful spectral features, e.g., a few Balmer lines, CH and CN lines, Mg triplet, lots of iron lines etc., for K giant identification as well as parametrization. Therefore, we use the S/N in $g$ band to quantify the quality of the spectra.}. Totally 2,046 matched objects are selected. 

In this sample, the \emph{true} K giant stars are defined as: \logg$<4$ when $4600<$\teff$<5600$ or \logg$<3.5$ when \teff$<4600$ (see the green polygon shown in figure~\ref{fig:SDSStefflogg}). It is noted that the SDSS SSPP pipeline does not fit M stars, i.e., \teff$<4000$, hence the SVM classifier trained by SDSS stellar parameters is also not suited for M giant stars. There are 274 \emph{true} K giant stars defined in the training dataset. The rest of the 2,046 stars in the sample are marked as non-K giant stars. The $EW$s of the spectral lines are measured from the corresponding LAMOST spectra of the training sample so that the signal to noise level, wavelength calibration, residual of sky subtraction etc. are comparable to the real dataset. 

We use the \emph{libsvm} package\footnote{http://www.csie.ntu.edu.tw/$\sim$cjlin/libsvm} in MATLAB to train the SVM classifier based on the training sample. 


\subsection{Validation and performance of the classifier}

Two different samples are selected as the test data to validate the classification. First, we use {the} MILES library, which provides high S/N low-resolution spectra. Then we test the performance using the LAMOST spectra combined with the \emph{true} K giant labels defined in SDSS stellar parameters. Table~\ref{tab:testsvm} summarizes the results of the performance tests.

\begin{figure}[htbp]
\begin{center}
\includegraphics[scale=0.6]{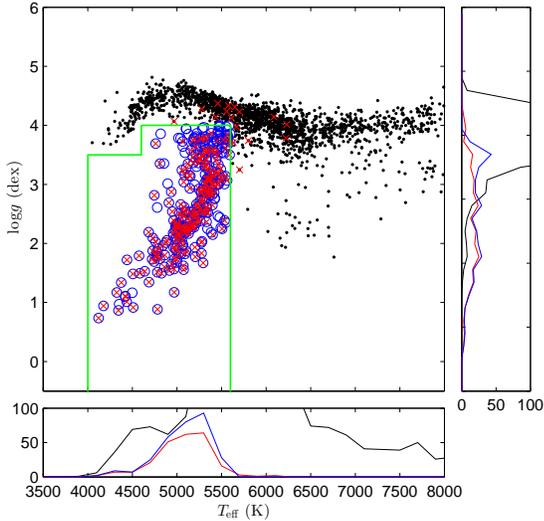}
\caption{{Distribution} in \teff\ vs. \logg\ for the LAMOST--SDSS cross-matched stars, for which \teff\ and \logg\ are from {the} SDSS SSPP pipeline. The blue circles are the \emph{true} K giant stars and the black dots the non-K giant stars. The red crosses indicate the \emph{identified} K giant stars from the SVM classifier. The right and bottom panels show the histograms of \logg\ and \teff, respectively. The black, blue, and red curves are the stellar counts for the full dataset, the \emph{true}, and the \emph{identified} K giant stars, respectively. The green rectangle shows the selection criteria of the \emph{true} K giant stars.}\label{fig:SDSStefflogg}
\end{center}
\end{figure}

\begin{figure}[htpb]
\begin{center}
\includegraphics[scale=0.6]{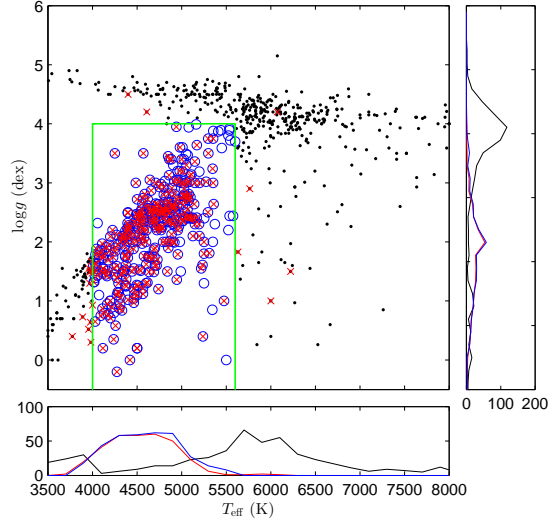}
\caption{Distribution in \teff\ vs. \logg\ for {the} MILES library. The symbols {and the sideway histograms are the same as} in Figure~\ref{fig:MILEStefflogg}.}\label{fig:MILEStefflogg}
\end{center}
\end{figure}

The MILES library contains 985 low-resolution spectra in total. Note that stars with \teff$>10000$\,K are not suitable for the Lick indexes because their Balmer lines are too broad. After eliminating these, 919 stars are left as the test dataset. There are 350 stars defined as the \emph{true} K giant stars with $4000<$\teff$<5600$\,K and \logg$<4$  (see the green rectangle shown in figure~\ref{fig:MILEStefflogg}). 
{It is noted that the K giant selection criteria for SDSS and MILES data are different. This is because that in SDSS parameter plane (see figure~\ref{fig:SDSStefflogg}) the \logg\ of the late-type main-sequence stars (\teff$<5000$\,K) decreases for unknown reason. Hence, the criteria of the K giant selection based on SDSS parameters avoid these dwarf stars. For the MILES case, as shown in figure~\ref{fig:MILEStefflogg} the main-sequence is quite normal in low-temperature end. Therefore, we simply use a rectangle to select the \emph{true} K giant stars. The different criteria may not lead to significant inconsistency, because the two \emph{true} K giant samples selected from the MILES data based on the two criteria are almost the same.}
The SVM classifier identifies 325 K giant stars, out of which 304 are \emph{true} K giant stars. The \emph{completeness} is defined as the ratio of the number of \emph{identified} to that of the total \emph{true} K giant samples, while the \emph{contamination} as the fraction of the non-K giant stars in the \emph{identified} K giant samples. Under these definitions, the completeness of the K giant stars identification for MILES library is 86.9\% and the contamination is only 6.5\%. The performance of the K giant selection is very good as shown in \teff--\logg\ diagram in figure~\ref{fig:MILEStefflogg}.

It is also noted that the MILES spectra have very high signal-to-noise ratio. In order to test the performance for lower quality data, such as the LAMOST spectra, artificial noise is added to the $EW$s to simulate the realistic situation. The classification algorithm is run 10 times with 10\% random Gaussian noise \footnote{It means that the sigma of the Gaussian is 10\% of the $EW$ of a line. It is equivalent with S/N=10.} added on the $EW$s and obtain the values of the mean completeness of 84.1\% and contamination of 6.7\%; when the noise level goes up to 20\% the mean completeness of 77.1\% and contamination of 8.3\% are obtained. Thus, we are convinced that the SVM classifier is robust even for low S/N spectra.

The second test is carried out using LAMOST regular survey data instead of MILES. We select 2,251 common stars in both LAMOST and SDSS DR9 catalog complying with the criteria in section~\ref{sect:trainsvm}. These test data are more representative than the MILES library, because they are observed and reduced in very similar situations to those for the training dataset. We define 302 \emph{true} K giant stars in the test sample using the same definition in section~\ref{sect:trainsvm}. We then apply the SVM classifier to them and classify 236 K giant stars with 220 \emph{true} K giant stars and 16 contaminates. In other words, the completeness of the classification is 72.8\% and the contamination is 6.8\%. Figure~\ref{fig:SDSStefflogg} shows the performance of the K giant selection in \teff--\logg\ diagram.

To investigate how the signal-to-noise ratio of the LAMOST spectra affect the classification, we separate the test data into two groups at S/N($g$)=20. For those with S/N($g$)$<20$, the completeness and the contamination are 72.1\% and 9.6\%, respectively, while the two values turn out to be 74.0\% and 2.2\% for those with S/N($g$)$>20$. 
Even though the LAMOST spectra are more affected by noise, the completeness of the test based on the LAMOST spectra with SDSS parameters is only $\sim10$\% lower than that of the first test based on MILES. The result is very promising in the sense that even in the very tough case, i.e., low S/N spectra, more than 70\% of the K giant stars can still be identified.

\begin{table*}
\begin{center}
\caption{The performance of the SVM K giant classifier.}\label{tab:testsvm}
\begin{tabular}{l|c|c}
\hline\hline
test data & completeness & contamination\\
\hline
MILES & 86.9\% &6.5\%\\
MILES (10\% noise)\tablenotemark{a} &84.1\% & 6.7\%\\
MILES (20\% noise) &77.1\% & 8.3\%\\
\hline
LAMOST+SSPP params.\tablenotemark{b} & 72.8\% & 6.8\%\\
LAMOST+SSPP params. (S/N($g$)$<$20) & 72.1\% & 9.6\%\\
LAMOST+SSPP params. (S/N($g$)$>$20) & 74.0\% & 2.2\%\\
\hline
LAMOST+LAMOST params.\tablenotemark{c} & 74.5\% &2.4\%\\
LAMOST+LAMOST params. (S/N($g$)$<$20)  & 66.7\% & 5.0\%\\
LAMOST+LAMOST params. (S/N($g$)$>$20)  & 79.3\% & 0.9\%\\
\hline\hline
\end{tabular}
\tablenotetext{a}{Add 10\% Gaussian noise to the $EW$s of the MILES spectra.}
\tablenotetext{b}{The test dataset contains the LAMOST spectra with SSPP derived stellar parameters.}
\tablenotetext{c}{The test dataset contains the LAMOST spectra with LAMOST derived stellar parameters.}
\end{center}
\end{table*}

\subsection{Application to the LAMOST survey}\label{sect:lamostkg}

\begin{figure}[htbp]
\begin{center}
\includegraphics[scale=0.7]{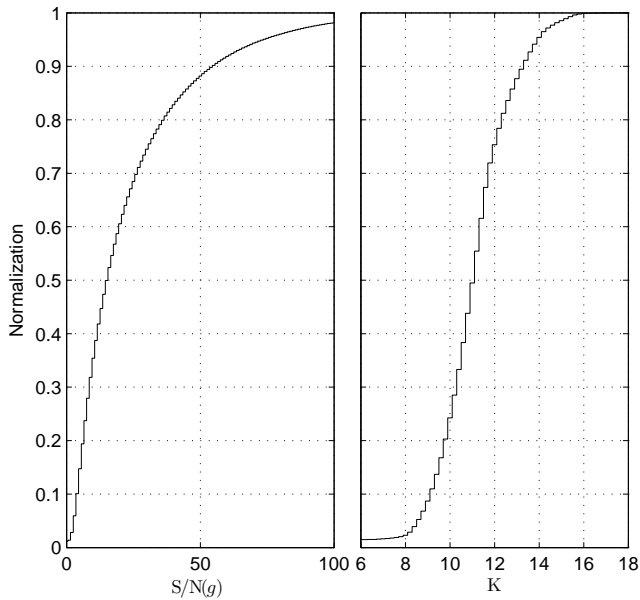}
\caption{\emph{Left panel: }The cumulative distribution of S/N($g$) for the 298,036 identified K giant stars. \emph{Right panel: }The cumulative distribution of 2MASS $K$ magnitude for the identified K giant stars.}\label{fig:kgiantKmag}
\end{center}
\end{figure}

We apply the SVM classifier to all $\sim$1.9 million stellar spectra from the LAMOST DR1 survey data. {The training process took about one day in a DELL workstation, while the process identification of K giant star in the whole dataset was very fast, taking about two hours with 12 parallel threads.} In total, we obtain 298,036  identified K giant spectra of which 196,440 (119,813) have S/N($g$)$>=10$ ($>=20$). The left panel of figure~\ref{fig:kgiantKmag} shows the distribution of the S/N($g$) for the identified K giant stars; and the right panel shows the distribution of 2MASS $K$ magnitude for the same sample. The median value of S/N($g$) is about 15 and the median $K$ magnitude for the samples is at 11\,mag.

\begin{figure}[htpb]
\begin{center}
\includegraphics[scale=0.6]{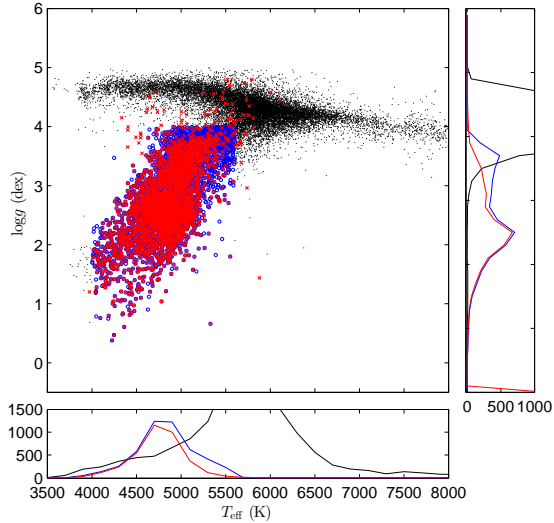}
\caption{Distribution in \teff\ vs. \logg\ of the LAMOST stars, for which \teff\ and \logg\ are given by LAMOST pipeline. The symbols are {the same} as in Figure~\ref{fig:MILEStefflogg}. In order to demonstrate the performance, only arbitrary one fiftieth spectra are drawn in the figure.}\label{fig:LMtefflogg}
\end{center}
\end{figure}

The LAMOST pipeline also provides stellar parameters for a selection of about 1 million high quality spectra. Among them, there are 238,332 \emph{true} K giant spectra based on the definition given in section~\ref{sect:trainsvm} but with stellar parameters provided by LAMOST pipeline. Applying the classification algorithm to the sample, the completeness and contamination becomes 74.5\% and 2.4\%, respectively, which is slightly better than the test with SDSS parameters (see Table~\ref{tab:testsvm}). When dividing the data into two groups at S/N($g$)=20, the completeness values of the low and high S/N data are 66.7\% and 79.3\%, contamination is 5.0\% and 0.9\%, respectively. Figure~\ref{fig:LMtefflogg} shows the selected K giant stars in \teff--\logg\ diagram.

It is clear that the classification algorithm is more successful for high quality data (high S/N); spectral quality of the survey is one of the keys to success.

\begin{figure*}[htbp]
\begin{center}
\includegraphics[scale=0.7]{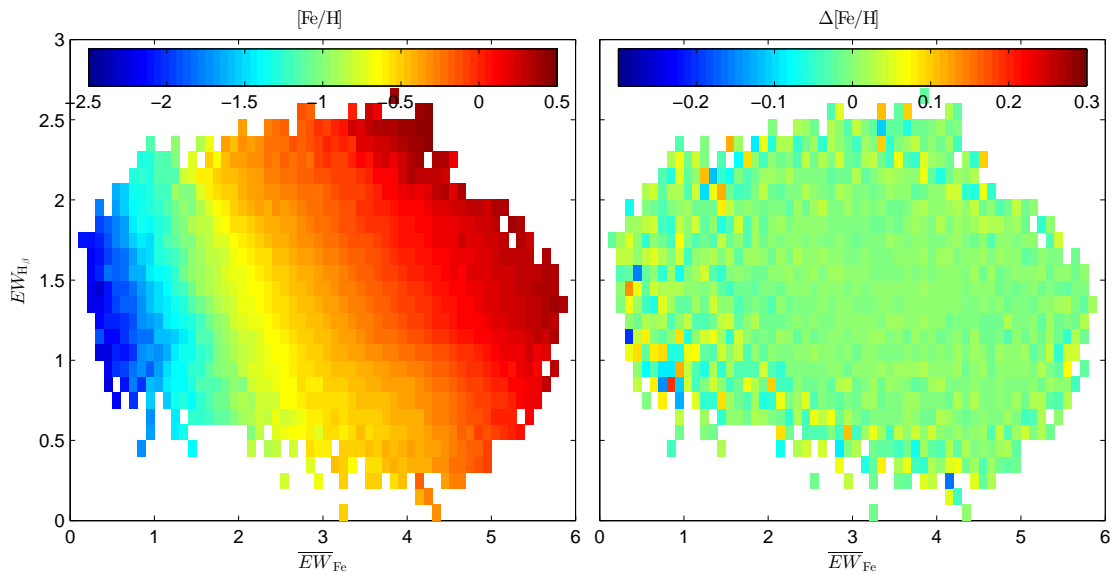}
\caption{ \emph{Left panel}: The {color codes }the median LAMOST \feh\} in \fe\ vs. \ewHb\ plane for 114,900 K giant stars with LAMOST \feh. \emph{Right panel}: The residual of the LM2D estimated \feh. }\label{fig:metalest}
\end{center}
\end{figure*}

\section{Metallicity estimation}\label{sect:metal}
In general, the distance of a giant star is a function of magnitude, color index, and metallicity. Therefore, the metallicity of K giant stars has to be determined before any reliable distance can be derived.

The LAMOST pipeline has provided metallicity estimates for about 1 million high quality spectra. However, there are lots of K giant stars, which are mostly low S/N data, selected in section~\ref{sect:lamostkg} with no estimated metallicity. In order to estimate the metallicity for all identified K giant stars including those with low S/N spectra, we use the LAMOST derived metallicity (denoted as \fehlm) as the training dataset and establish an estimator (hereafter named LM2D) to estimate \feh\ from the line features.

\begin{figure}[htbp]
\begin{center}
\includegraphics[scale=0.55]{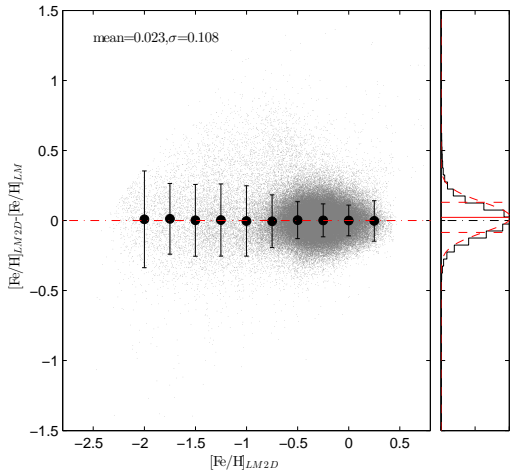}
\caption{The residual of the \feh\ estimation from LM2D compared with the LAMOST \feh. The gray dots stand for the individual stars. The big black dots with error bars are the median residual at each \feh\ bin. The distribution of the residual is shown in the right side {panel} as the black line. The red dashed curve shows the best Gaussian fit of the distribution and the solid and dashed horizontal lines show the mean and 1-$\sigma$ values, respectively.}\label{fig:metalestperf1}
\end{center}
\end{figure}

In general, the metallicity of the K giant stars is a function of both the Fe lines and the effective temperature. Therefore, we simultaneously select \ewHb, which is correlated with \teff, and \fe, which is the mean $EW$s of 9 Fe lines at 4383$\rm\AA$, 4531$\rm\AA$, 4668$\rm\AA$, 5015$\rm\AA$, 5270$\rm\AA$, 5335$\rm\AA$, 5406$\rm\AA$, 5709$\rm\AA$, and 5782$\rm\AA$, to estimate \feh.

The training dataset is selected with the following criteria: i) they are labeled as K giant stars based the LAMOST \teff\ and \logg\ values according to section~\ref{sect:trainsvm}; ii) they are also identified as K giant stars in the SVM classifier; iii) they have S/N($g$)$>$20; iv) they have \feh\ measured by LAMOST; and v) \ewHb$>$0. There are 114,900 K giant stars that meet all of these conditions. 

The {median} \feh\ of the sampled stars forms a curved surface in \ewHb\ vs. \fe\ plane (see the left panel of figure~\ref{fig:metalest}), which can be fitted with a 2-dimensional thin-plate spline function\footnote{The thin-plate spline is a generalized spline used for two or more dimensional interpolation and smoothing.} depending on both \ewHb\ and \fe. The best fit thin-plate spline is then used to predict the metallicity of the K giant stars.
The right panel of figure~\ref{fig:metalest} shows the residuals of the spline fitting for the 114,900 K giant stars. For most of the area the residual is smaller than 0.1\,dex.  Figure~\ref{fig:metalestperf1} shows \fehlmd-\fehlm\ as a function of \fehlm\ for the training dataset. Although the dispersion of the residual of \feh\ becomes larger at low metallicitiy end, no significant systematics is found in figure~\ref{fig:metalestperf1}. The overall dispersion of the differential \feh\ is 0.11\,dex, being smaller for metal-rich stars but larger for metal-poor stars. The very weak line features in metal-poor spectra are responsible for the poor measurements of the equivalent widths and consequently account for {the} larger uncertainty in the metallicity estimation.

\begin{figure}[htbp]
\includegraphics[scale=0.55]{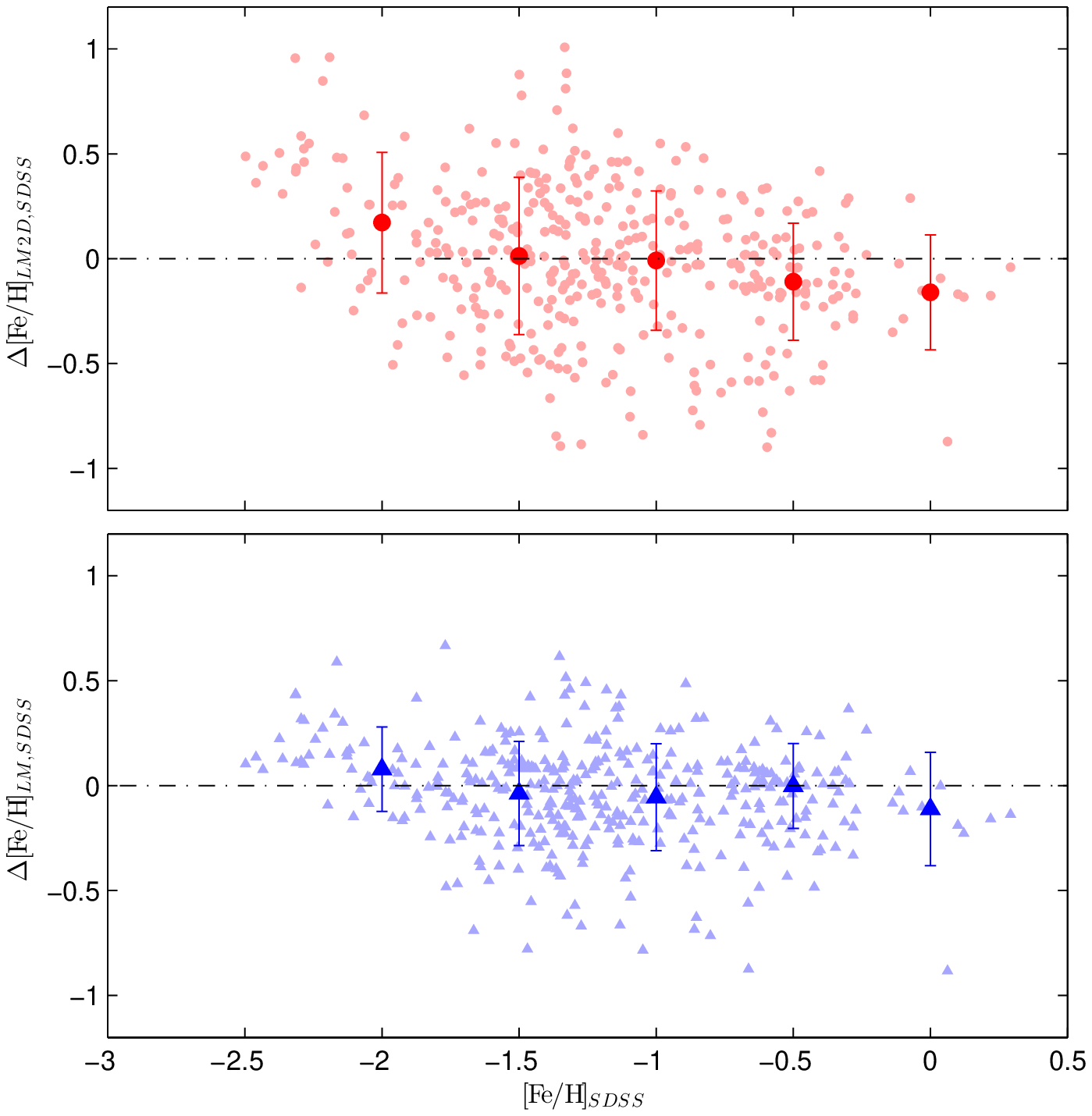}
\caption{\emph{Top panel: }The difference between \fehlmd\ and \feh$_{SDSS}$ as a function of \feh$_{SDSS}$ for 394 K giant stars with LM2D, LAMOST, and SDSS metallicities (pale red dots). The filled red circles with error bars show the medians and standard deviations at \feh=-2, -1.5, -1, -0.5, and 0. \emph{Bottom panel: } The difference between \fehlm\ and \feh$_{SDSS}$ as a function of \feh$_{SDSS}$ (pale blue triangles). The blue triangles with error bars are the {medians and standard deviations}.}\label{fig:fehfeh}
\end{figure}

\begin{figure}[htbp]
\includegraphics[scale=0.55]{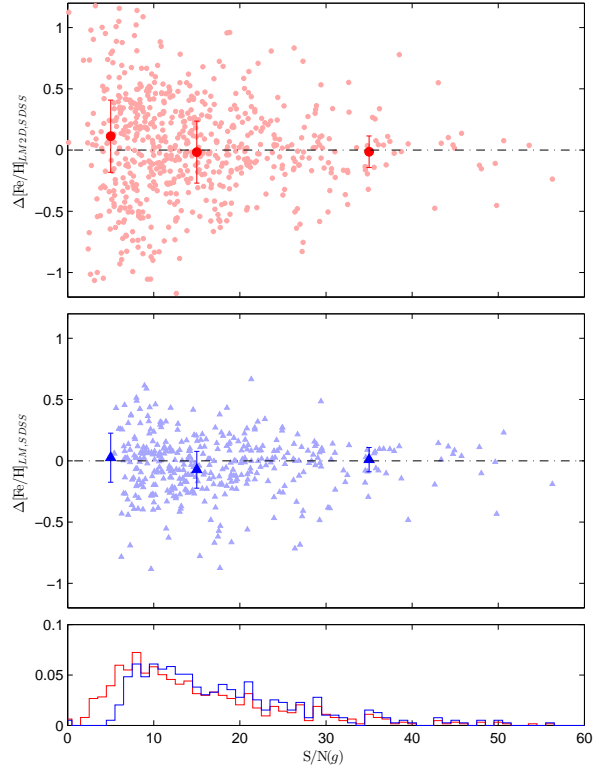}
\caption{\emph{Top panel: }The difference between \fehlmd\ and \feh$_{SDSS}$ as a function of $g$-band signal-to-noise ratio, S/N($g$), in LAMOST spectra for 636 K giant stars with both LM2D and SDSS metallicities (pale red dots). The filled red circles with error bars show the medians and the median absolute deviations at S/N($g$)=0-10, 10-20, and $>20$. \emph{Middle panel: } The difference between \fehlm\ and \feh$_{SDSS}$ as a function of S/N($g$) for 394 K giant stars with both LAMOST and SDSS metallicities (pale blue triangles). The blue triangles with error bars are the median and standard deviation values. \emph{Bottom panel: }The red (blue) line shows the histogram of the S/N for the K giant stars with LM2D (LAMOST) metallicity.}\label{fig:snrfeh}
\end{figure}

The comparisons of the derived metallicity \fehlmd\ with \fehlm\ and SDSS metallicity, \feh$_{SDSS}$, will provide more tests of the performance of the LM2D method.
We start by collecting the test data sample.  The LAMOST dataset is cross-matched with SDSS DR9 and cleaned using the following criteria: 1) the S/N of the SDSS spectra is higher than 20; 2) The stellar parameters are provided by SDSS; 3) they are K giant stars according to criteria defined in section~\ref{sect:trainsvm} using the SDSS parameters, and 4) they are also identified as K giant stars with the SVM classifier. A sample of 636 K giant stars with \fehlmd\ are resolved, of which 394 also have LAMOST \fehlm. Notice that the SDSS data are dominated by metal-poor stars due to their fainter magnitude limit along with the fact that they are at higher latitude(see figure~\ref{fig:fehfeh}). Hence, the uncertainty of the LM2D metallicity is larger than the average level{,} as expected from figure~\ref{fig:metalestperf1}.

Figure~\ref{fig:fehfeh} shows the comparison between \fehlmd\ (\fehlm) and \feh$_{SDSS}$ using the 394 stars with both \fehlmd\ and \fehlm.
In the top panel,  it is indicated that the LM2D overestimates the \feh\ for stars with \feh$_{SDSS}<$-1 by at about 0.2\,dex and underestimates the value by the similar level at \feh$_{SDSS}=0$. As a comparison, the bottom panel shows that the LAMOST \feh\ has similar trend with smaller bias and dispersion.

\begin{table}[htbp]
\begin{center}
\caption{The performance of the \fehlm\ and \fehlmd\ compared with SDSS metallicity.}\label{tab:fehperf}
\begin{tabular}{c|c|c|c}
\hline\hline
S/N($g$) & 0-10 & 10-20 & $>20$ \\
\hline
$\Delta$[Fe/H]$_{LM,SDSS}$\tablenotemark{a} & 0.03 & -0.07 & 0.01 \\
$\sigma_{LM,SDSS}$\tablenotemark{b} & 0.20 & 0.15 & 0.10\\
\hline
$\Delta$[Fe/H]$_{LM2D,SDSS}$\tablenotemark{c} & 0.11 & -0.02 & -0.01\\
$\sigma_{LM2D,SDSS}$\tablenotemark{d} & 0.29 & 0.25 & 0.13\\
\hline
$\Delta$[Fe/H]$_{LM2D,LM}$\tablenotemark{e} & 0.00 & 0.04 & -0.02\\
$\sigma_{LM2D,LM}$\tablenotemark{f} & 0.28 & 0.20 & 0.10\\
\hline\hline
\end{tabular}
\tablenotetext{a}{$\Delta$[Fe/H]$_{LM,SDSS}=$\fehlm-\feh$_{SDSS}$}
\tablenotetext{b}{The median absolute \\deviation of \fehlm-\feh$_{SDSS}$.}
\tablenotetext{c}{$\Delta$[Fe/H]$_{LM2D,SDSS}=$\fehlmd-\feh$_{SDSS}$.}
\tablenotetext{d}{The median absolute \\deviation of \fehlmd-\feh$_{SDSS}$.}
\tablenotetext{e}{$\Delta$[Fe/H]$_{LM2D,LM}=$\fehlmd-\fehlm.}
\tablenotetext{f}{The median absolute \\deviation of \fehlmd-\fehlm.}
\end{center}
\end{table}

Figure~\ref{fig:snrfeh} demonstrates the influence of the S/N of the LAMOST spectra in the LM2D metallicity estimation. The top panel presents the difference between \fehlmd\ and \feh$_{SDSS}$ as a function of S/N($g$). Not surprisingly, the dispersion of the difference of metallicity declines when the signal-to-noise ratio increases. The middle panel shows the same trend in LAMOST \fehlm, although the dispersions are smaller than those in \fehlmd. As shown in the bottom panel, it is clear that LM2D can be applied to less preferable data, i.e., S/N($g$) can be as low as 3, though the measured metallicity will be somewhat less accurate.

Table~\ref{tab:fehperf} shows the offsets and dispersion of \fehlm\ (\fehlmd) compared with \feh$_{SDSS}$ as a function of S/N($g$). 
 Compared with SDSS parameters, the uncertainty of LM2D method for K giant stars with S/N($g$)$<10$ is only about 0.1\,dex larger than that of the LAMOST metallicity. And the uncertainty of LM2D is similar as that of LAMOST metallicity for the spectra with S/N($g$)$>$20. 

\section{The estimation of distance}\label{sect:dist}
The estimation of the distance for K giant stars is a non-trivial task, because the K giant stars are distributed along the red giant branch, therefore the luminosity is a steep function of effective temperature or color index.

\citet{xue12} estimated the distance with accuracy down to 0.2\,mag in distance moduli (DM) using cluster-based fiducial isochrones. Although the clusters provide more realistic isochrones than synthetic data and hence match the observed data, the 4 clusters and one BaSTI {\citep{basti}}isochrone (at \feh=0) they used are only sparsely located at 5 discrete values in \feh. The estimated distance is very sensitive to the accuracy of the \feh, particularly for metal-poor stars. Thus, a denser grid in \feh\ may improve the accuracy of distance.

In this section, we estimate the distance of the halo K giant stars in an alternative way, i.e., using synthetic isochrones with calibration based on globular clusters. Because the disk K giant stars have much broader range of age, their distance estimation needs different method and beyond this work.

In principle, the synthetic isochrones can provide arbitrary grid in \feh. In our case, the mean uncertainty of the metallicity is $\sim$0.3\,dex. Therefore, we use the synthetic library with $\Delta Z=0.0001$, which corresponds to the  $\Delta$[Fe/H]$\sim$0.3\,dex at \feh={-2}\,dex. To correct the discrepancy between the synthetic isochrones and the real observed objects, we calibrate the synthetic data using globular clusters.

\subsection{Method}\label{sect:distmethod}

Because there is no unified accurate photometry covering a large range of $r=10$--18\,mag, which matches the LAMOST spectra, the photometry of the targets are composed of different catalogs. UCAC4 is used for the targeting brighter than $r=14$\,mag, while SDSS and PanSTARRS are used for the targeting fainter than 14\,mag. To avoid additional work of transforming different photometry systems to an unified one for all LAMOST spectra, we use 2MASS \citep{cutri03}, which is in near infrared bands covering most of the magnitude range but with less accuracy, as a compromising solution to provide a homogeneous distance estimation for both bright and faint stars. 

We use a dense grid of synthetic isochrones \citep{girardi02,marigo08} at a typical age of 10\,Gyr, which is suitable for the halo stars.  The library is tailored to contain only RGB+SGB tracks. The posterior probability density function (PDF) of the absolute magnitude in $K$ band, $M_K$, for a giant star can be obtained from the following equation according to Bayes' theorem:
\begin{align}\label{eq:postpdf}
p&(M_K|(J-K)_0, {\rm [Fe/H]}, {\rm isochrone})=\nonumber \\
p&((J-K)_0, {\rm [Fe/H]}|M_K, {\rm isochrone})p(M_K),
\end{align} 
where the term $p((J-K)_0, {\rm [Fe/H]}|M_K, {\rm isochrone})$ is the likelihood and the prior $p(M_K)$ the luminosity function. {$(J-K)_0$ is the reddening corrected color index. We adopt the extinction map from \citet{sfd98} and the extinction parameters for $J$ and $K$ bands from \citet{ccm89} given $R_V=3.1$.} In this work, we adopt the luminosity function at 10\,Gyr from the library website \footnote{http://stev.oapd.inaf.it/cgi-bin/cmd}, which is based on Chabrier IMF \citep{chabrier01}. Note that $(J-K)_0$ is estimated from photometric observation and \feh\ is derived from spectroscopic survey data, therefore, they are supposed to be two independent measurements. Then the likelihood can be separated into:
\begin{align}\label{eq:likeli}
p&((J-K)_0, {\rm[Fe/H]}|M_K, {\rm isochrone})=\nonumber\\
p&((J-K)_0|M_K, {\rm isochrone})p({\rm[Fe/H]}|M_K, {\rm isochrone}).
\end{align}
The two terms on the right hand side of {equation}~(\ref{eq:likeli}) can be specified as the following equations providing Gaussian errors for the measurement of $J-K$ and \feh.

\begin{align}\label{eq:likeligauss1}
p&((J-K)_0|M_K,{\rm isochrone})\sim \nonumber\\
&exp(-{((J-K)_{iso}-(J-K)_0)^2\over{2\sigma_{J-K}^2}}),
\end{align}
\begin{align}\label{eq:likeligauss2}
p&({\rm[Fe/H]}|M_K,{\rm isochrone})\sim \nonumber \\
&exp(-{({\rm[Fe/H]}_{iso}-{\rm[Fe/H]})^2\over{2\sigma_{{\rm[Fe/H]}}^2}}),
\end{align}
where the variables in {equations}~(\ref{eq:likeligauss1}) and (\ref{eq:likeligauss2}) with subscript \emph{iso} are quantities from isochrone and those without subscript are from the observed data of a given star. The $\sigma_{J-K}$ and $\sigma_{\rm[Fe/H]}$ are measurement uncertainties of $(J-K)_0$ and \feh\ for the star.

The posterior PDF of the absolute magnitude for a star derived from {equation}~(\ref{eq:postpdf}) is then converted to that of DM by applying apparent $K$ band magnitude. {In the next sections we use the median value of DM as the best estimated value. The $-\sigma$ and $+\sigma$ of DM are defined by the 15\% and 85\% percentiles of the PDF.} 
 
\subsection{Calibration with globular clusters}
\begin{figure}[htbp]
\begin{center}
\includegraphics[scale=0.5]{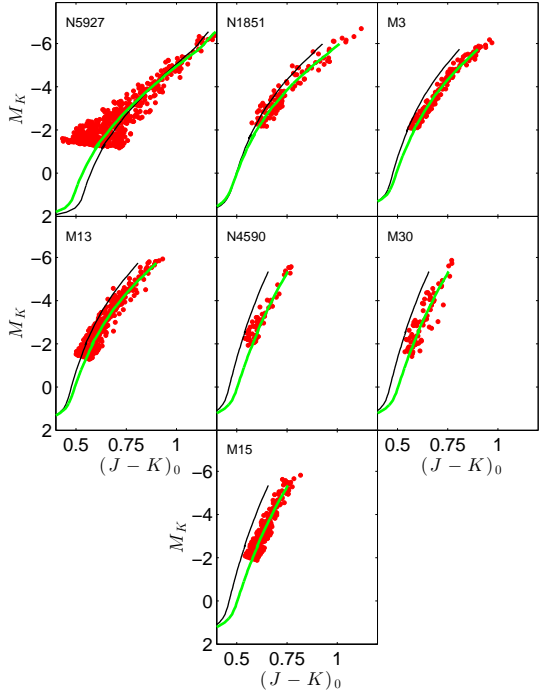}
\caption{The black lines are the synthetic isochrones with the same \feh\ as the globular clusters. The thick green lines are the calibrated isochrones according to {equation}~(\ref{eq:calibrateDJK}). The red dots are the K giant members of the globular clusters.}\label{fig:GCcalibrate}
\end{center}
\end{figure}

\begin{figure*}[htbp]
\begin{center}
\includegraphics[scale=0.55]{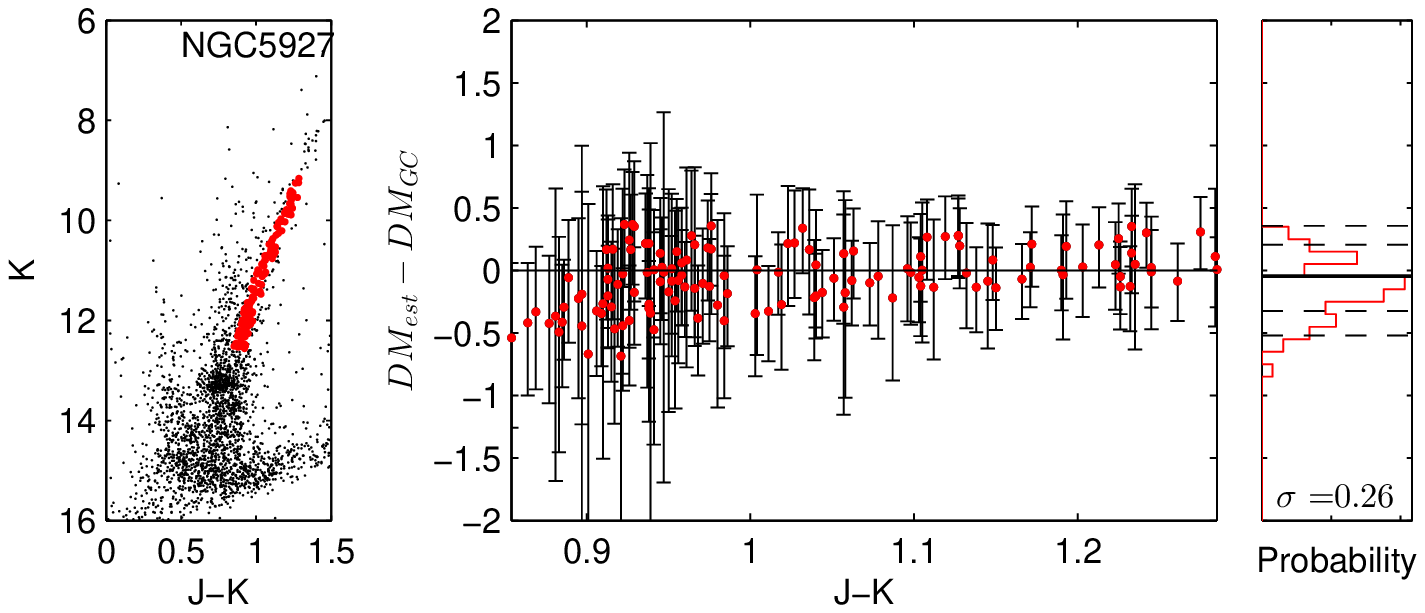}
\includegraphics[scale=0.55]{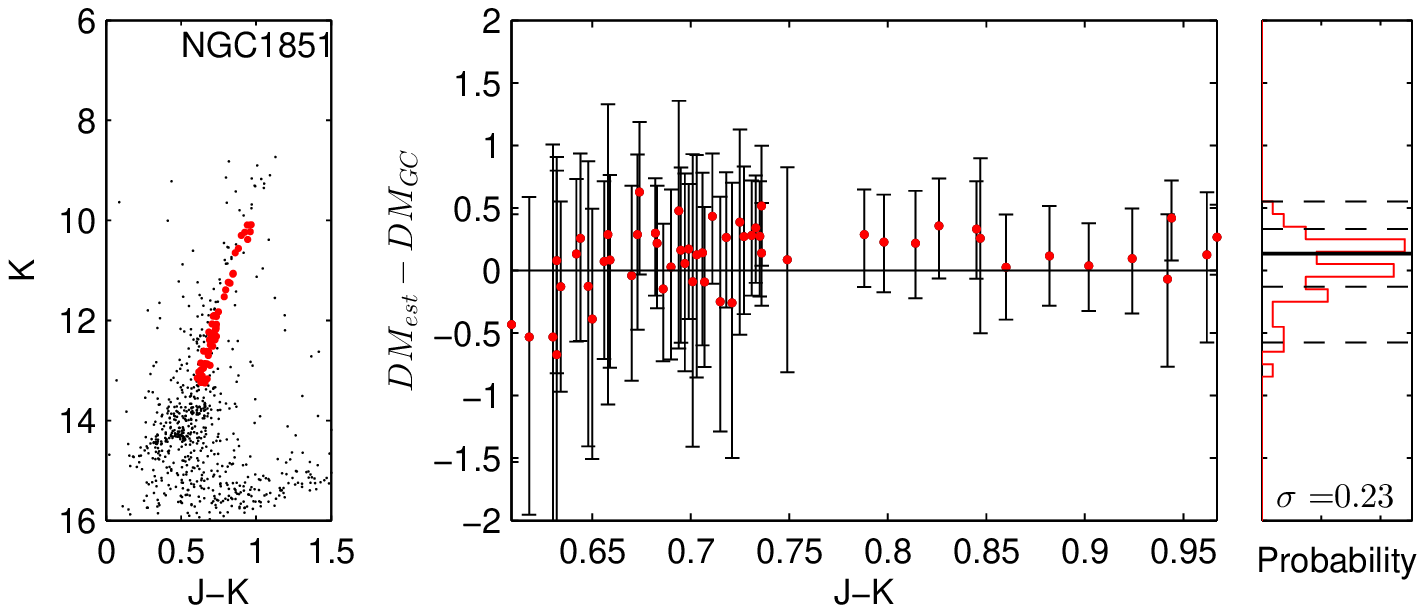}
\includegraphics[scale=0.55]{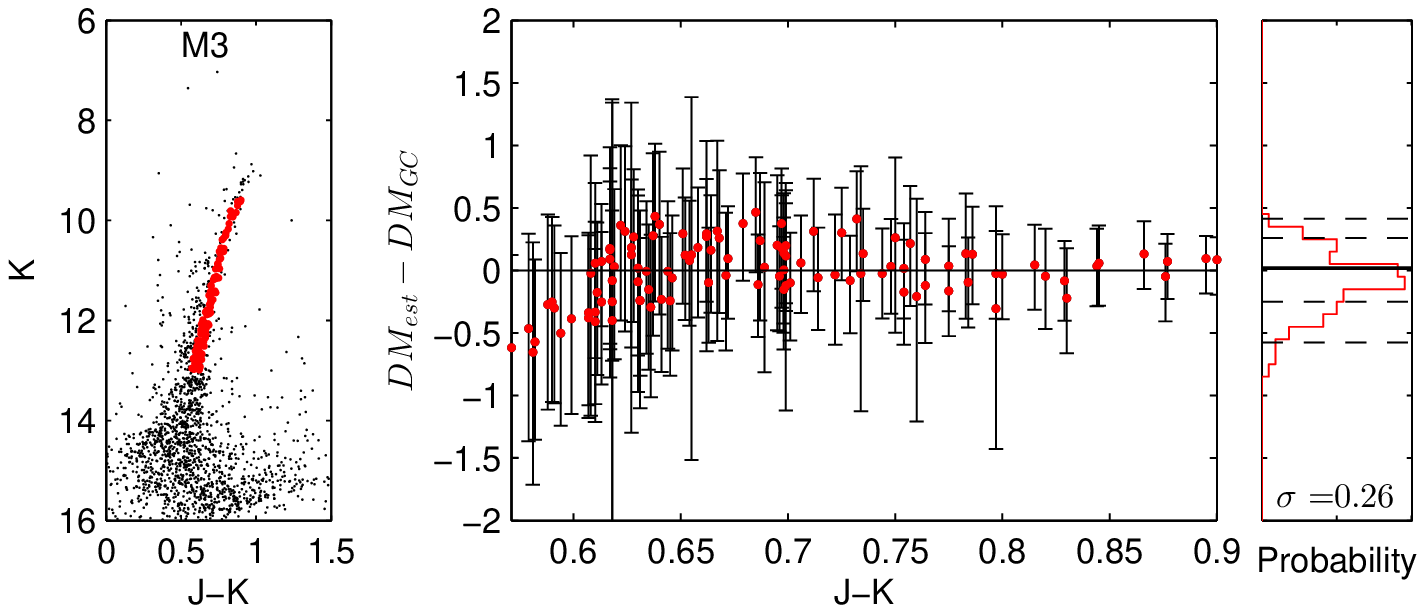}
\includegraphics[scale=0.55]{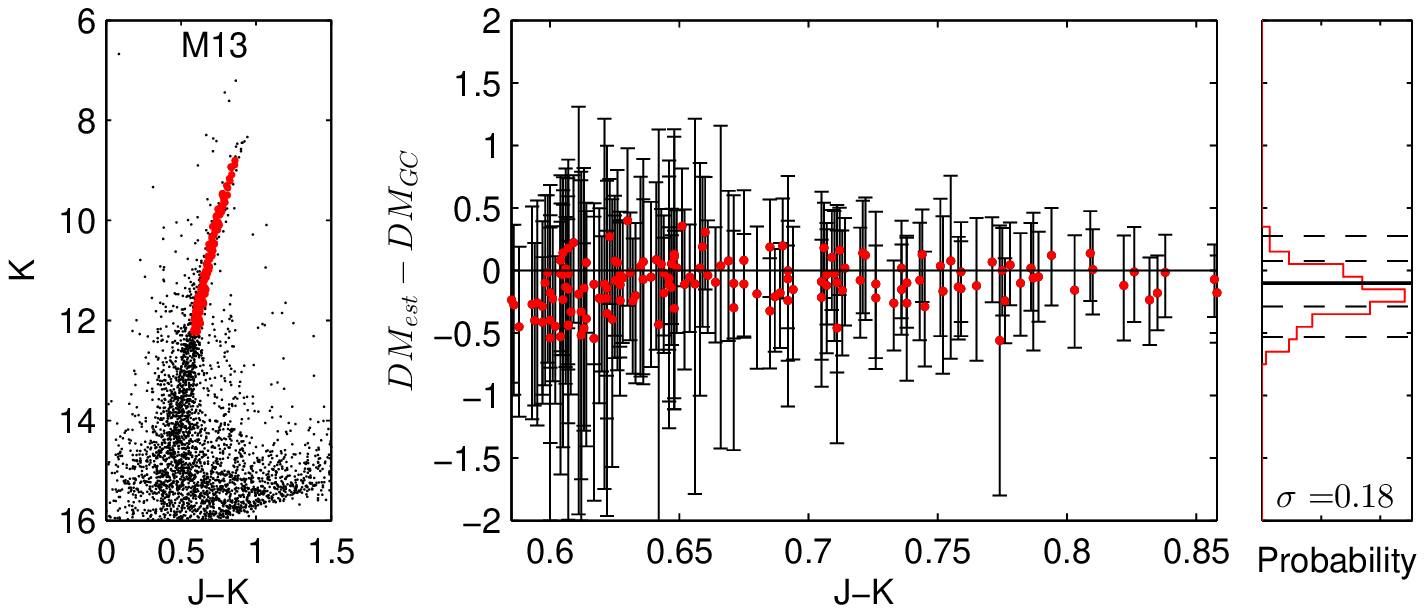}
\includegraphics[scale=0.55]{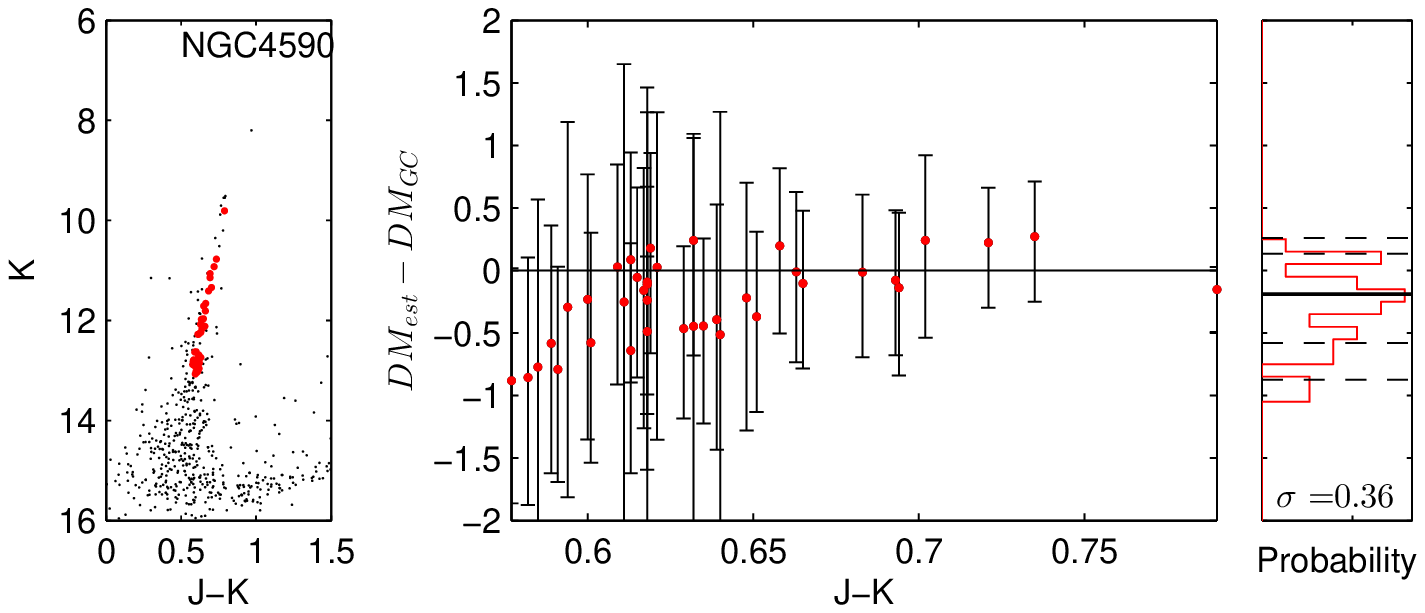}
\includegraphics[scale=0.55]{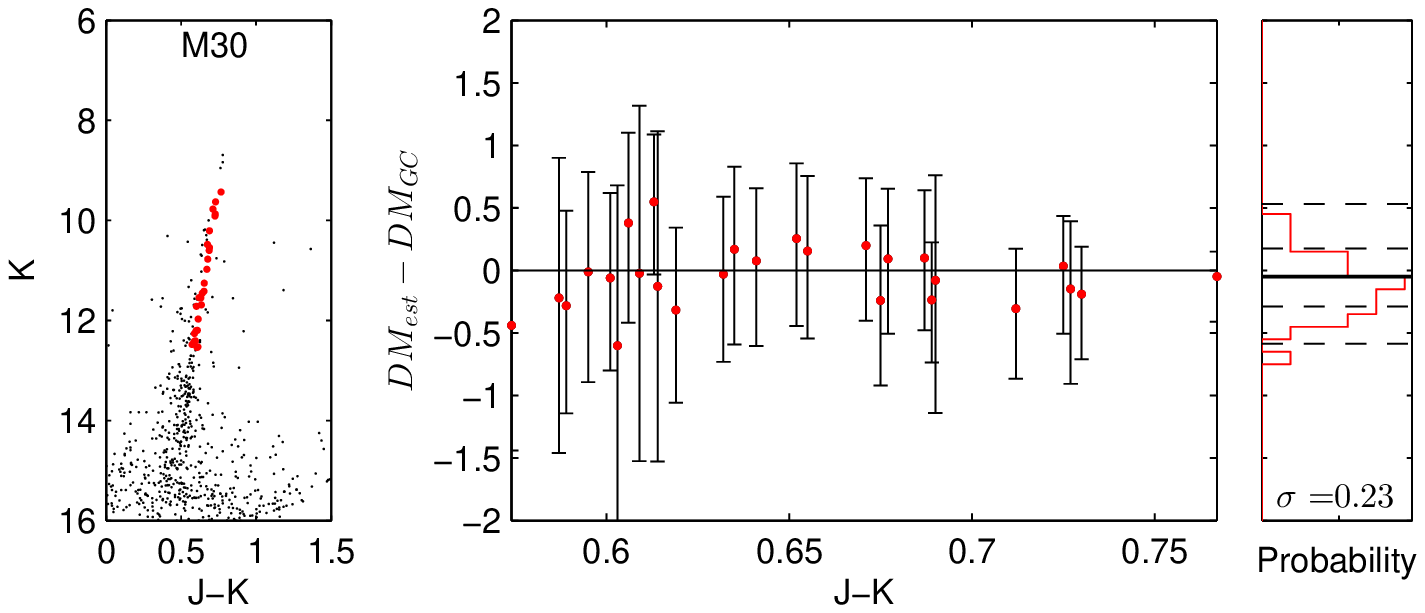}
\includegraphics[scale=0.55]{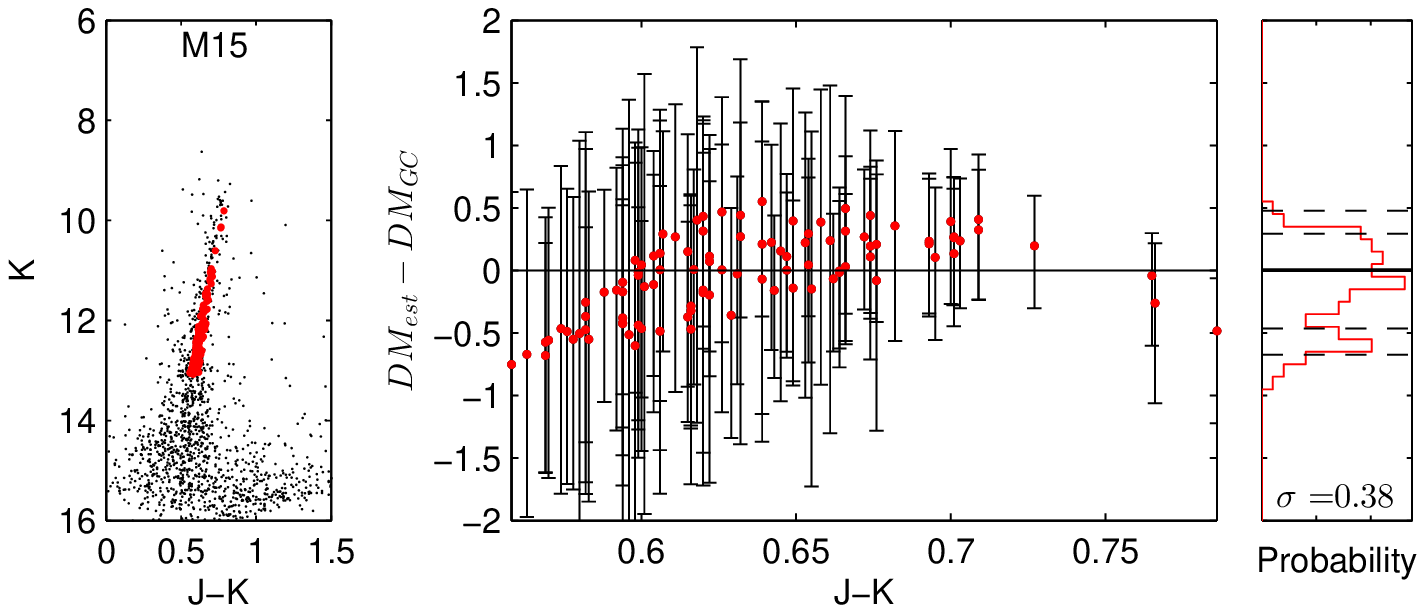}
\caption{The distance performance tested with globular clusters NGC5927, NGC1851, M3, M13, NGC4590, M30, and M15. For each panel, the left plot shows the color-magnitude diagram for the stars around the cluster (black dots) and the selected K giant members (red dots). The middle one shows the residuals of the distance moduli of the K giant members (red dots) with 1-$\sigma$ uncertainties as a function of $J-K$. The right one shows the distribution of the residual DM (red lines). The solid and dashed horizontal lines indicate the peak, 1-$\sigma$, and 2-$\sigma$ values.}\label{fig:GC1}
\end{center}
\end{figure*}

Distance estimation using synthetic isochrones, as described above, needs to be calibrated observationally. A usual practice is to use globular clusters (GCs). The main issue is that the isochrones in the observed data are different from that of the synthetic spectra, given the same metallicity, and the offset becomes larger for metal-poor objects (see figure~\ref{fig:GCcalibrate}). 
We select 7 GCs (see table~\ref{tab:gc}), which are bright and cover a wide range in metallicity, to calibrate the color offsets. The giant members of the GCs are manually selected from the $(J-K)_0$\,vs.~$K$ diagram of stars within a small radius (defined in the 5th column of table~\ref{tab:gc}) around the center of each cluster (the red dots shown in the color-magnitude diagrams in figures~\ref{fig:GCcalibrate} and \ref{fig:GC1}).
Note that although they are not definite GC members, the contaminations do not affect the result since we replace the individual metallicities for the selected stars with the mean value of the GC in the calibration and later test.
 
\begin{table}[htbp]
\begin{center}
\caption{The parameters of 7 GCs for calibration}\label{tab:gc}
\begin{tabular}{c|c|c|c|c}
\hline\hline
GC names & Distance\tablenotemark{a} & \feh\tablenotemark{a} & E(B-V)\tablenotemark{b} & selection \\&&&&radius\tablenotemark{c}  \\
\hline
&(kpc)&dex&mag&degree\\
\hline
NGC5927 & 7.7 & -0.49 &  0.45 & 0.1 \\
NGC1851 & 12.1 & -1.18 & 0.02 & 0.07 \\
M3 & 10.2 & -1.5 & 0.02 & 0.2 \\
M13 &  7.1 & -1.53 & 0.02 & 0.32 \\
NGC4590 & 10.3 & -2.23 & 0.05 & 0.07 \\
M30 &  8.1 & -2.27 & 0.03 & 0.07 \\
M15 & 10.4 & -2.37 & 0.1 & 0.32 \\
\hline
\hline
\end{tabular}
\tablenotetext{a}{\citet[][revision 2010]{harris}}
\tablenotetext{b}{{Color excess in $B-V$, adopting the values from \\\citet[][revision 2010]{harris}}}
\tablenotetext{c}{The radius used to select the giant members around the \\center of each GC.}
\end{center}
\end{table}

We apply a polynomial surface model to obtain the corrected offset color index, $\Delta(J-K)$, as a function of $(J-K)_0$ and \feh. The best fit is

\begin{align}\label{eq:calibrateDJK}
\Delta(J-K)=&-0.1421\pm0.0187+ \nonumber\\
&0.08454\pm0.0209(J-K)_0- \nonumber\\
&0.06482\pm0.0170{\rm[Fe/H]}- \nonumber\\
&0.09195\pm0.0166(J-K)_0{\rm[Fe/H]}- \nonumber\\
&0.0194\pm0.0026{\rm[Fe/H]}^2.
\end{align}  

Figure~\ref{fig:GCcalibrate} demonstrates the performance of the calibration. The red dots are the K giant members of the 7 globular clusters, the black lines show the synthetic isochrones with same metallicity as the corresponding GCs. After employing the calibration of $(J-K)_0$ using {equation}~(\ref{eq:calibrateDJK}), the isochrones are shifted to the green lines, well fit the GC data.

Adding $\Delta(J-K)$ into {equation}~(\ref{eq:likeligauss1}), we obtain the new likelihood of $(J-K)_0$:
\begin{align}\label{eq:likelicalibrate}
p&((J-K)_0|M_K,{\rm isochrone})\sim \nonumber \\
&exp(-{((J-K)_{iso}+\Delta(J-K)-(J-K)_0)^2\over{2\sigma_{J-K}^2}}).
\end{align}

\begin{figure}[htbp]
\begin{center}
\includegraphics[scale=0.5]{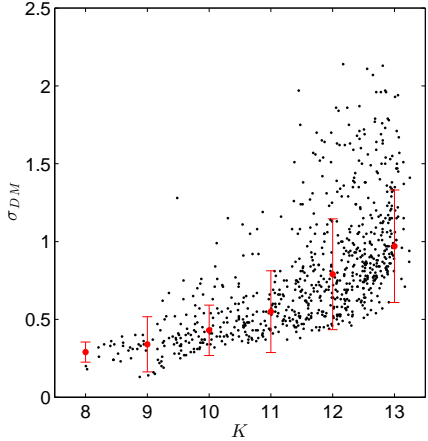}
\caption{The black dots are the error of DM, $\sigma_{DM}$, of the giant members of the 7 globular clusters as a function of apparent magnitude $K$. The red dots present the median values at given $K$ magnitude.}\label{fig:DistPerf}
\end{center}
\end{figure}

The distances of the K giant members are derived by using the above calibration. Figure~\ref{fig:GC1} shows the performance of the estimation using the member giant stars of the 7 GCs. Figure~\ref{fig:DistPerf} shows the errors of DM (denoted as $\sigma_{DM}$) as a function of apparent magnitude $K$. At 11\,mag, the median value of the full K giant sample, the mean error of DM is about 0.6\,mag, or $\sim$30\% in distance.
This is a factor of 2.5 larger than that of \citet{xue12}. The use of 2MASS, which is in infrared, shallower, and less accurate than SDSS, is likely responsible for the poorer performance. The distance measurement is inevitably compromised until a unified, high accuracy optical photometric catalog is established for the LAMOST survey.

\subsection{Measuring distance for LAMOST K giant stars}
\begin{figure}[htbp]
\begin{center}
\includegraphics[scale=0.55]{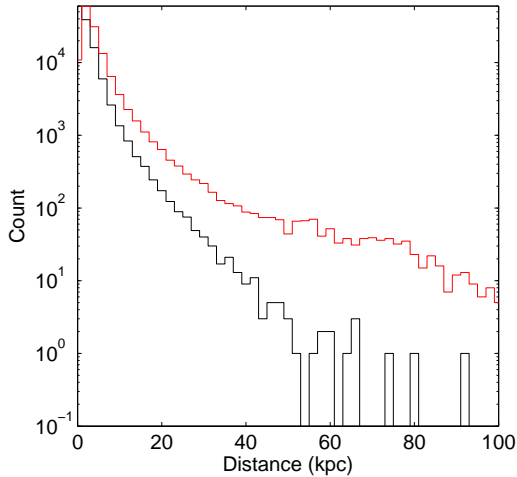}
\caption{The distribution of the distance for the K giant stars. The red and black lines represent for the K giant stars selected from the SVM classifier and from the LAMOST derived parameters, respectively.}\label{fig:3Ddist}
\end{center}
\end{figure}

Before deriving distance of the K giant stars, the sample is purified by the following procedure: i)  only the later/latest epoch observation is kept for duplicated objects; ii) the non-stellar targets are removed;
iii) only the stars with $5<K<15.5$\,mag and the $K$ magnitude error lower than 0.1\,mag are selected; iv) the stars with $|b|<10^\circ$ are removed to avoid the high extinction and the majority of the disk K giant stars; v) the K giant stars with \fehlmd$<-3$\,dex, which may not be reliable, are also excluded; and vi) the red clump stars selected according to Appendix A are excluded, since the distance estimation method described in section~\ref{sect:dist} is not suitable for them. After these cuts, there are 134,597 RGB/SGB-like K giant remaining.

Meanwhile, we clip the \emph{true} K giant samples selected from LAMOST parameter catalog (see subsection~\ref{sect:lamostkg}) using the similar criteria mentioned in previous paragraph, and obtain 76,822 \emph{true} K giant stars left for distance estimation with \fehlm.

In figure~\ref{fig:3Ddist}, the spatial distribution of the K giant stars based on \fehlmd\ (red) and LAMOST \fehlm\ (black) are plotted. The star count of LM2D at 20\,kpc is a factor of 4 larger than that of LAMOST sample at the same distance. The factor increases to around 10 when the distance is larger than 40\,kpc. Therefore, our identification method significantly increases the K giant sampling of the survey, which is more efficient in larger distances, allowing the search for the kinematic substructure in a much larger volume.

\subsection{Caveats of the distance estimation}\label{sect:caveats}

\begin{figure}[htbp]
\begin{center}
\includegraphics[scale=0.65]{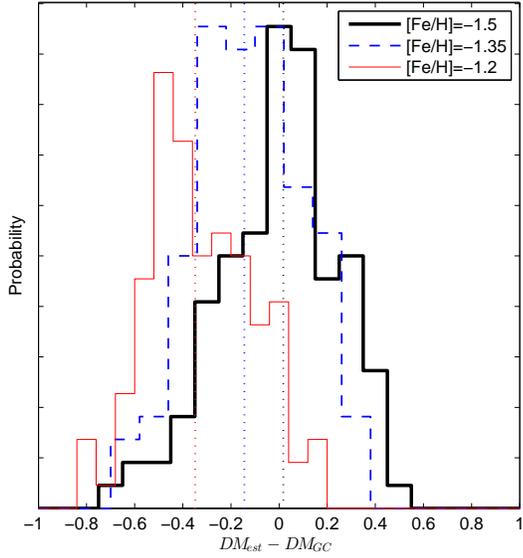}
\caption{The effect of the overestimation of the metallicity in the estimated distance. The histograms show the residual distribution of the distance moduli of the M3 member K giant stars when the applied metallicity is -1.5 (the true metallicity of the globular cluster, shown in thick black line), -1.35 (blue dashed line), and -1.2 (red thin line), respectively. Their median residuals are marked using vertical black, blue, and red dot lines, respectively.}\label{fig:distsystematic}
\end{center}
\end{figure}

\begin{figure}[htbp]
\begin{center}
\includegraphics[scale=0.65]{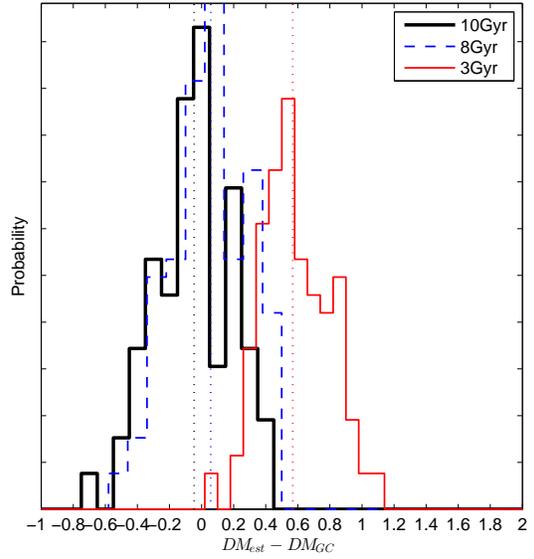}
\caption{The effect of the age in the estimated distance. The histograms show the residual distribution of the distance moduli of the NGC5927 member K giant stars when the applied age of the isochrones is 10 (thick black line), 8 (blue dashed line), and 3\,Gyr (red thin line), respectively. Their median residuals are marked using vertical black, blue, and red dot lines, respectively.}\label{fig:distsystematicage}
\end{center}
\end{figure}

Because the accuracy of distance estimation depends on the accuracy of the metallicity, the systematic bias of the estimated metallicity, as shown in figure~\ref{fig:fehfeh}, may induce a systematic bias in estimated distance. Figure~\ref{fig:distsystematic} shows the distribution of the residual DM for the K giant members of M3. With the true metallicity, i.e., -1.5 dex according to table~\ref{tab:gc}, the estimated distance does not show any significant systematic bias. However, when the metallicity of the member stars is overestimated by 0.15 (0.3)\,dex, i.e., \feh=-1.35 (-1.2\,dex), the median residual DM shifts 0.15 (0.35)\,mag towards left, which equivalent with 7\% (17\%) underestimation of the distance.

For the halo and thick disk K giant stars, the distance estimation based on isochrones with age of 10\,Gyr should be acceptable. However, for younger stars in the disk this may lead to a systematic bias in the distance estimation. Figure~\ref{fig:distsystematicage} shows that for the metal-rich globular cluster NGC5927, which true metallicity is -0.49, the residual of the DM shifts by about 0.6\,mag toward right when a set of 3\,Gyr isochrones are applied in the distance estimation. In other words, the distance estimation method may underestimate by $\sim30$\% for this case.

As a summary, the distance may be underestimated either for metal-poor stars since both the LAMOST pipeline and the LM2D method tend to slightly overestimate their metallicity or for young disk stars due to the application of relatively old isochrones.

\section{The Sagittarius stream}
\label{sect:sgr}

In order to assess and demonstrate the validity of our catalog, we
carry out the following scientific verification. The Milky Way's
stellar halo is known to host many tidal streams and substructures,
which can be identified using catalogs such as the one presented
here. The most prominent stream originates from the Sagittarius dwarf
galaxy \citep{ibata01}. This stream has been detected by a number
of authors using a variety of datasets, both photometric and
spectroscopic  \citep[][etc.]{newberg02,majewski03,belokurov06,dohmpalmer01,majewski04a,monaco07,carlin12}. These detections can then be
used to probe the potential of the Milky Way \citep[e.g.,][]{helmi01, law05, fellhauer06, law10} 
 or to estimate the properties of
the host galaxy, which was originally much more massive than it is
today \citep{niedersteostholt10}.

To aid our search for Sagittarius members, we convert the equatorial
coordinates into the \citet{belokurov14} system \citep[itself based on the
system of][]{majewski03}, described by two angles
$\tilde{\Lambda}_\odot$ and $\tilde{B}_\odot$. This system has the
stream located along the equator, with the core at
$\tilde{\Lambda}_\odot = 0$ and this angle increasing in the direction
of motion, i.e. the start of the leading stream lies at positive
$\tilde{\Lambda}_\odot$ and the trailing stream at negative
$\tilde{\Lambda}_\odot$. An illustration of this coordinate system is
shown in figure 2 of \citet{belokurov14}, where it can be seen
that the stream lies in the region $|\tilde{B}_\odot| \lesssim 10$ deg.
We correct for extinction using the maps of \citet{sfd98}.

\begin{figure}
\begin{center}
\plotone{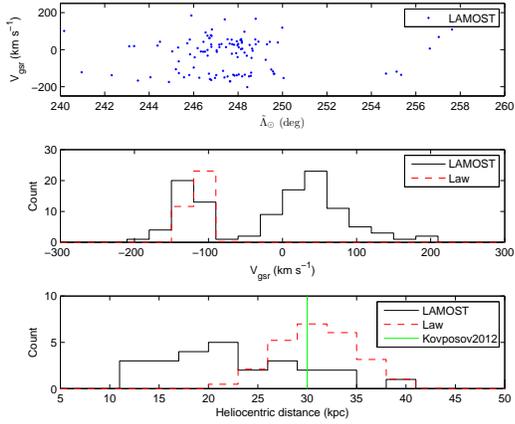}
\end{center}
\caption{The detection of Sagittarius trailing debris. The top panel
shows the velocity signature of the stream in our K giant sample
around $|\tilde{B}_\odot| < 15$ deg. The detection is even more evident
when we plot a histogram of Galacto-centric radial velocities
(${\rm V_{gsr}}$) for these stars (middle panel), as can be seen from
the peak around $-160$ to $-100$ km/s. The dashed line shows the
prediction from the model of \citet{law10}, where we have only
included stars stripped in the past 3 Gyr. The bottom panel shows the
distances of these giant stars with velocities consistent with the
stream (i.e. $-160<{\rm V_{gsr}}<-100$ km/s), along with the
prediction from Law's model. There is a small ($\sim20$ per
cent) offset, which is discussed in Section \ref{sect:sgr}.} 
\label{fig:sgr_south}
\end{figure}

We begin by searching for stream members in the South Galactic cap
region, where one can easily detect material from the trailing
arm. This has been studied in detail and its properties well known
\citep[see][for an overview]{koposov12}. The distances at this
location are around 30\,kpc and the velocities significantly offset
from the background halo at around $-150$ to $-100$ km/s. This can be
seen clearly in the upper panel of figure~\ref{fig:sgr_south}, with the
stream members separated from the bulk of the halo population. In the
middle panel we present a histogram of radial velocities for all stars
with $|\tilde{B}_\odot| < 15$ deg and, for comparison, show the
distribution of particles in this region from the model of \citet{law10}, where we have only included stars which have been stripped
in the past 3 Gyr. There is good agreement here, which is unsurprising
as the model was tuned to match the velocity signal in this part of
the stream, but it is reassuring that our K giant sample is able to
pick up a clean sample of Sagittarius members. The lower panel shows the
distances for these stars, retaining only those which have $-160 <
v_{gsr} < -100$ km/s. Here we can see that there is a slight offset {in distance}
between our data and Law's model, at a level of around 20 per
cent. Again it is known that the Law model does not have problems in
this region, as evidenced by good agreement with main-sequence
turn-off stars from \citet{koposov12}. Therefore this offset is
most likely due to systematics in our distances, which we discuss below.

Another avenue to verify the catalogue is to look for material at the
trailing tail's apo-centre. This was recently detected by \citet{belokurov14} and \citet{drake13} in both blue horizontal branch
stars and red giants and lies at a distance of around 100
kpc. Although the radial velocity is, by definition, close to zero in
this region, the signal should be relatively strong due to the small
number of smooth halo stars this far out in the halo.

\begin{figure}
\begin{center}
\plotone{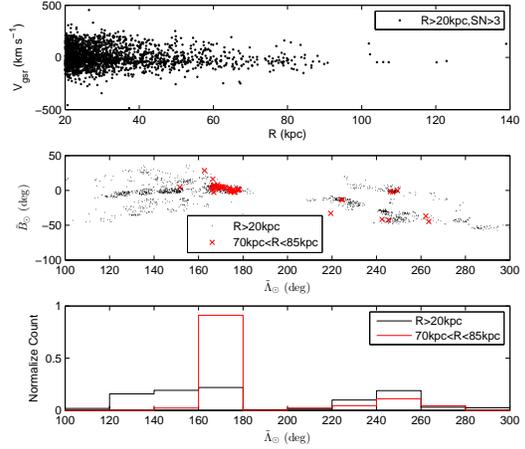}
\end{center}
\caption{The detection of distant Sagittarius debris, corresponding to
the apo-centre of the trailing tail. The top panel
shows the distribution of galacto-centric radial velocity (${\rm
V_{gsr}}$) and galacto-centric distance (R) for all stars with
signal-to-noise larger than 3. There appears to be a clump of stars
around 80kpc. The spatial distribution of these distant stars ($70 <
{\rm R} < 85$ kpc; red crosses) is compared with the underlying
distribution (${\rm R} > 20$ kpc; black points). {The middle panel shows that these stars are also concentrated in a small range of $\tilde{\Lambda}_\odot$ between 160$^\circ$ and 180$^\circ$.} The bottom panel
shows the histograms of $\tilde{\Lambda}_\odot$ for these two
distributions and, again, it is clear that the distant giants are not drawn
uniformly from the underlying distribution.}
\label{fig:sgr_apo1}
\end{figure}

This indeed proves to be the case, as can be seen in
figure~\ref{fig:sgr_apo1}. The top panel shows how the radial velocity
varies as a function of Galacto-centric distance for all stars in our
catalogue and there is already a hint of a detection in the clump of
stars around 80 kpc. The nature of these stars is uncovered in the
middle panel, where we show their location on the sky. They are
clearly not drawn uniformly from the underlying population, but are
obviously clumped around $\tilde{B}_\odot \approx 0$ deg and
$\tilde{\Lambda}_\odot \approx 170$ deg, which is precisely
where the Belokurov detection lies. This finding is reinforced in the
lower-panel, which shows the distribution of these stars as a function
of $\tilde{\Lambda}_\odot$ - their distribution is dramatically
different from the general footprint of the LAMOST survey and so this
finding cannot be explained by a quirk of the survey strategy.

\begin{figure}
\begin{center}
\plotone{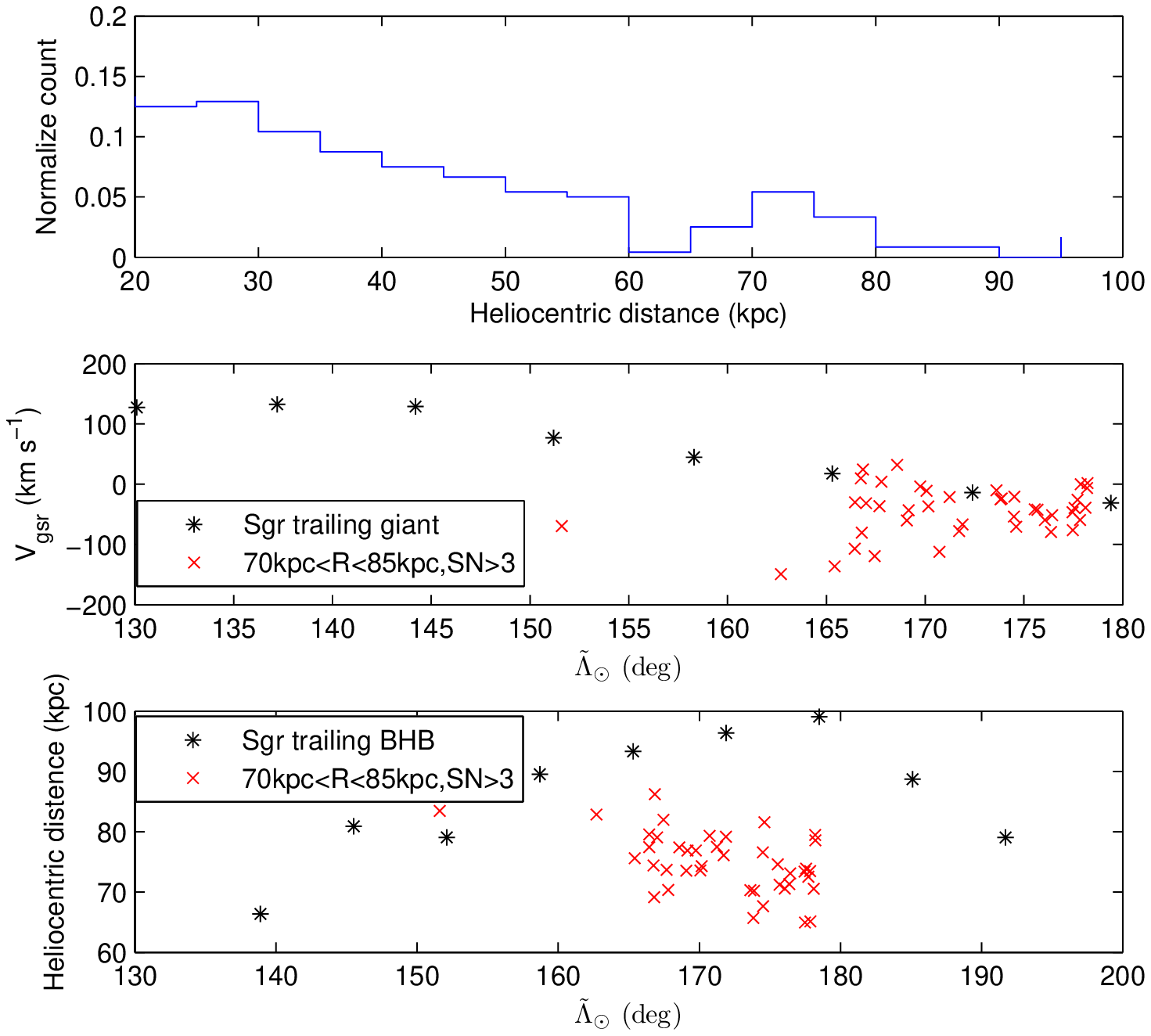}
\end{center}
\caption{A comparison between our detection of the distant Sagittarius
debris and that of \citet{belokurov14}. The top panel shows 
the distribution of helio-centric distances for our LAMOST giants (with
${\rm S/N} > 3$) in the region where these
distant stars lie ($|\tilde{B}_\odot| < 30$ deg, $160 <
\tilde{\Lambda}_\odot < 180$ deg). A clear overdensity is visible at
helio-centric distances of around 65 to 80 kpc, corresponding to
Galacto-centric distances $70 < {\rm R} < 85$ kpc. The middle panel
compares our LAMOST giant velocities (red crosses) to Belokurov's SDSS
giant velocities (black asterisks). The same notation is used in the
lower panel, where we compare our giant distances to their blue
horizontal branch star distances. There is a small ($\sim20$ per cent)
offset, which is discussed in Section \ref{sect:sgr}.}
\label{fig:sgr_apo2}
\end{figure}

We compare our detection to that of \citet{belokurov14} in
figure~\ref{fig:sgr_apo2}. The top panel shows the distribution of helio-centric distances for our K giants in the region of Belokurov's apo-centre
detection ($|\tilde{B}_\odot| < 30$ deg, $160 < \tilde{\Lambda}_\odot
< 180$ deg). This figure demonstrates that the bulk of these stars
are in good agreement with their velocities (middle panel), although
our distances again appear to be systematically offset when compared
to Belokurov's blue horizontal branch star distances (bottom
panel). As before, this distance offset is at around the 20 per cent
level.

To summarize, we have shown that our catalogue is well-suited to the
detection of halo substructures and we believe that it will be a rich
resource for finding further streams. Such endeavors are beyond the
scope of this work, but we are now pursuing this goal and will report
our findings in a future publication. However, our distance estimates
do need to be handled with care. Systematic offsets could be due to
various factors, such as incorrectly estimated metallicities or ages
(Section \ref{sect:caveats}), or possibly contamination from red clump
stars (Appendix \ref{sect:appendix}). 




\newpage

\section{Conclusions}

We have established a SVM classifier directly from the spectra features, and then apply it to LAMOST spectra for K giant star selection. The method does not depend on the stellar parameters, e.g., \teff\ and \logg, thus has a broader range of capability to work on spectra with S/N as low as 3.  Tested with SDSS, MILES, and LAMOST data, the SVM classifier can select K giant stars from the survey dataset with 70-80\% completeness and a few percent contamination.
From the DR1 released $\sim$1.9 million stellar spectra, we identified about 290,000 K giant stars. Consequently, we expect that, when the survey will be concluded in five years, there will be a factor of 4 more K giant stars to be observed.

In order to estimate the distance of the K giant stars, we have firstly estimated the metallicity using LM2D method. Comparisons with SDSS parameters indicates that the total error of the estimation is between 0.1\,dex and 0.3\,dex. The advantage of the estimation method is that we can provide metallicity for the identified K giant stars with signal-to-noise ratio down to 3. 

We have then developed a Bayesian method to estimate the distance of the K giant stars using 2MASS photometry and the estimated metallicity from LM2D. The synthetic isochrone-based method is calibrated with 7 globular clusters. Therefore, the systematic bias due to the discrepancy between the synthetic and observed data is corrected. The uncertainty of the distance estimation is investigated using the same globular clusters, which covers a wide range in metallicities. We conclude that the uncertainty in DM is around 0.5\,mag at $K\sim11$\,mag, corresponding to about 30\% in distance.   

Given the distance and radial velocities of the K giant stars selected from the survey, we successfully identified many candidate members of the Sgr stream. These identifications demonstrate that there may be thousands of  K giant members of such tidal substructures to be observed and identified over a broad area of the sky by the end of the five-year survey. Consequently, it will be one of the most important homogeneous spectroscopic dataset to map the kinematics as well as the chemical abundance of them. This will provide a tight constraint on the dark matter mass in the Galactic halo as well as the forming history of the substructure themselves. 


\acknowledgements
This work is supported by the Strategic Priority Research Program "The Emergence of Cosmological Structures" of the Chinese Academy of Sciences, Grant No. XDB09000000 and the National Key Basic Research Program of China 2014CB845700. CL acknowledge the National Science Foundation of China under grants 11373032 and U1231119.
JLC and HJN acknowledge National Science Foundation under grant AST 09-37523.   
XXX acknowledges the Alexandra Von Humboldt foundation for a fellowship and the National Natural Science
Foundation of China under grants 11103031, 11233004 and 11003017.

Guoshoujing Telescope (the Large Sky Area Multi-Object Fiber Spectroscopic Telescope LAMOST) is a National Major Scientific Project built by the Chinese Academy of Sciences. Funding for the project has been provided by the National Development and Reform Commission. LAMOST is operated and managed by the National Astronomical Observatories, Chinese Academy of Sciences.

\appendix

\section{Remove the red clump stars}\label{sect:appendix}
Even in low extinction regions, there are quite a lot red clump stars in the K giant sample. The distance estimation method is not suitable for this type of stars since the isochrone is only for RGB/SGB stars. Hence, we have to identify and remove them from the samples so that we can use a purified dataset in the study of the spatial distribution of the K giant stars.

\begin{figure}[h]
\begin{center}
\begin{minipage}{17cm}
\centering
\includegraphics[scale=0.65]{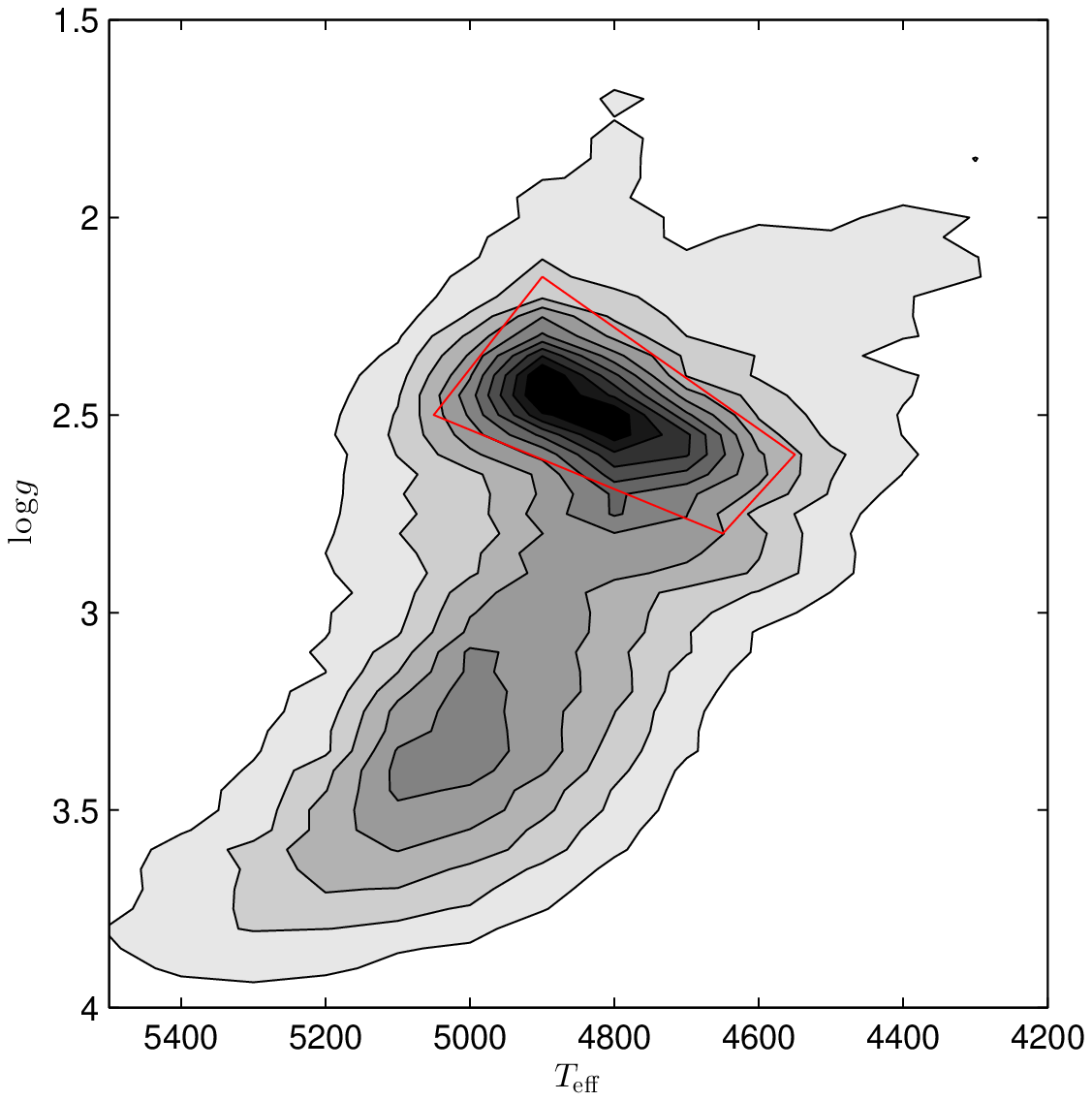}
\includegraphics[scale=0.65]{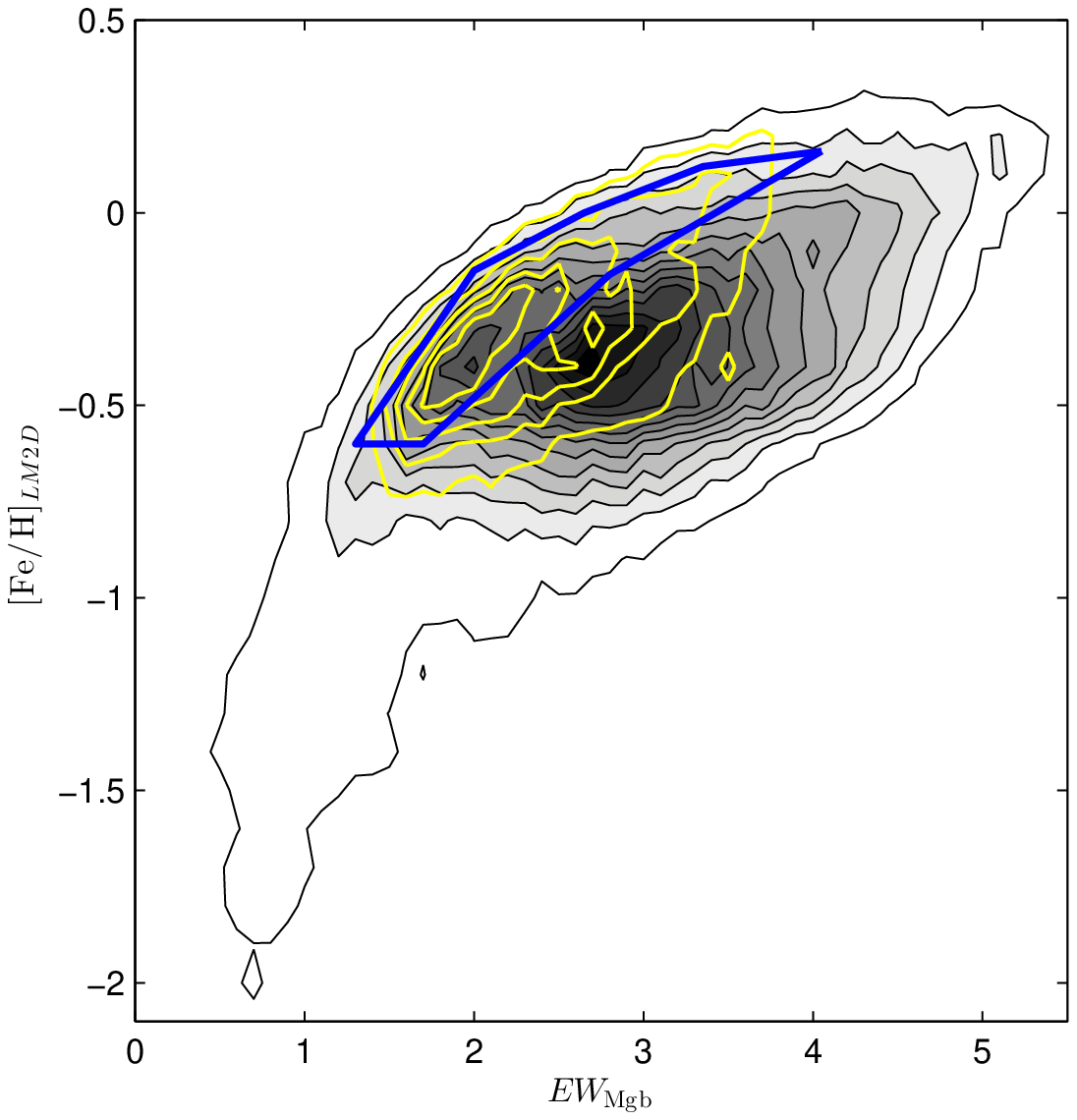}
\includegraphics[scale=0.65]{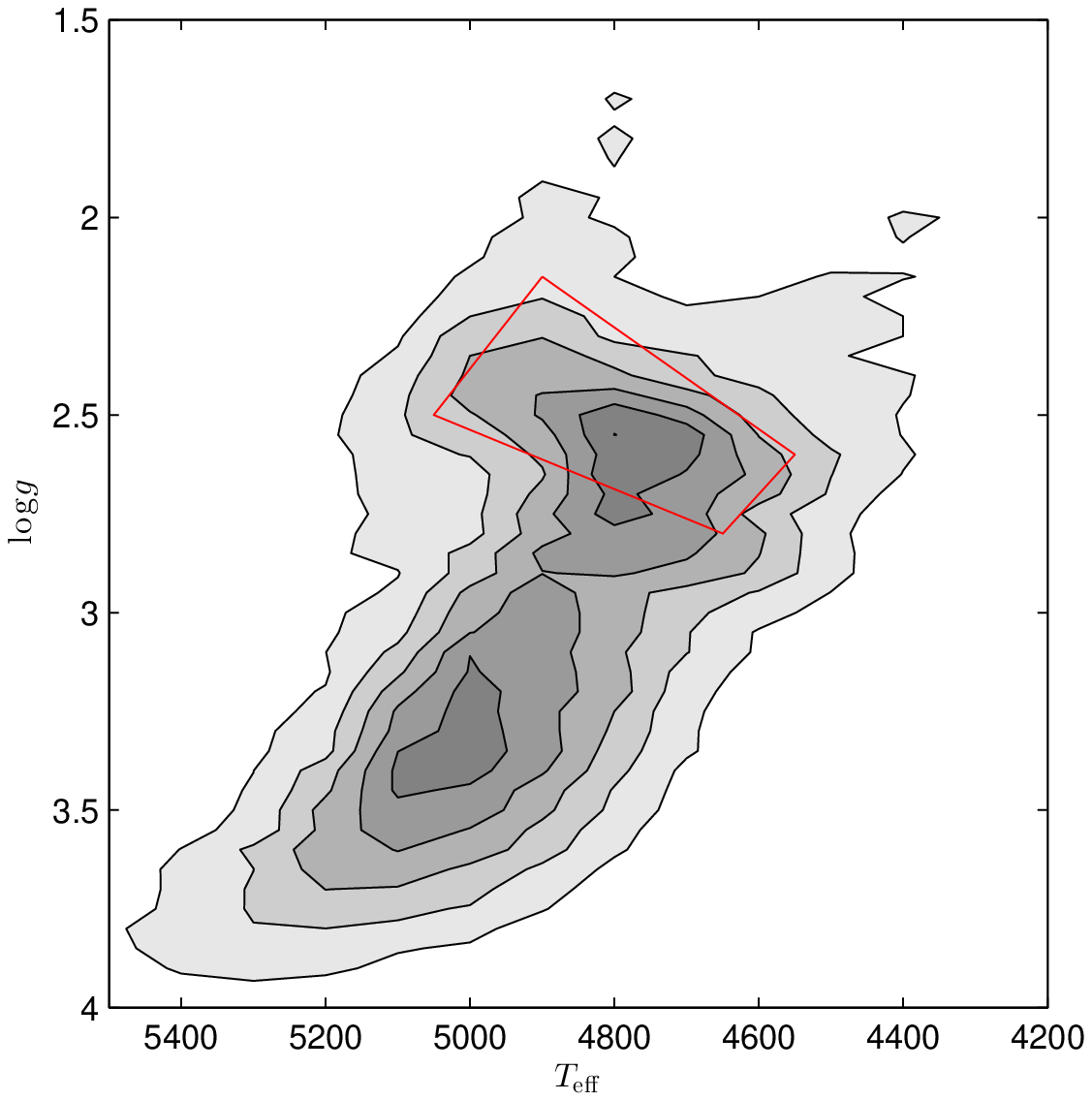}
\includegraphics[scale=0.65]{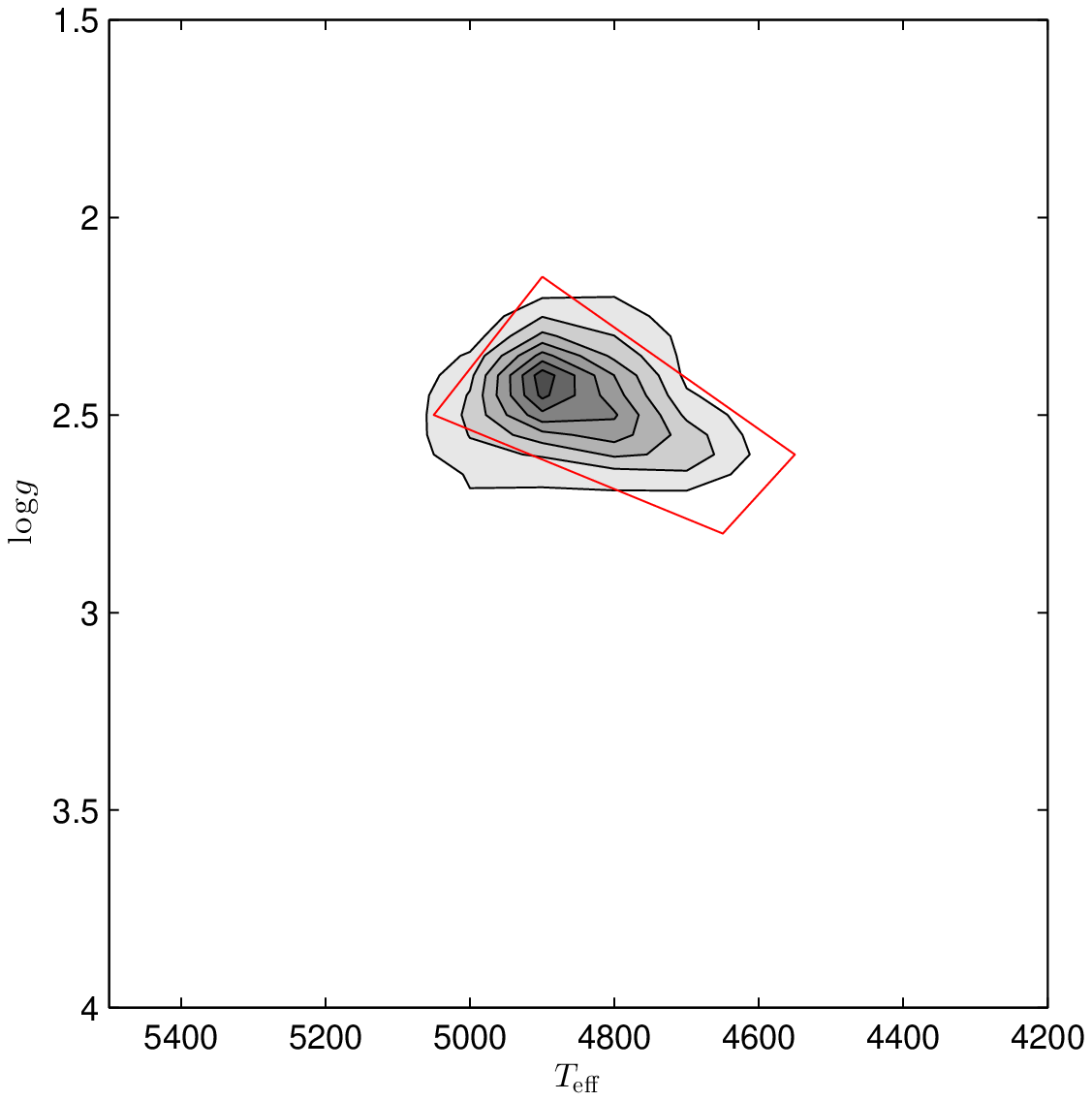}
\caption{\emph{Top-left panel}: The contour shows the distribution of the stars with LAMOST stellar parameters in \teff\ vs. \logg. The red polygon indicates the location of the red clump stars. 
\emph{Top-right panel}: The K giant stars (black filled contours) in $EW_{Mg_b}$\ vs. \fehlmd. The overlapped yellow contours are the stars located in the red polygon. The levels of the both contours are the star counts per unit $EW_{Mg_b}\times$\fehlmd\ varying from 5000 to 80000 with step size of 5000. The thick blue polygon is the simple selection criteria to define the red clump stars in $EW_{Mg_{b}}$ vs. \fehlmd. \emph{Bottom-left panel}: The \teff\ vs. \logg\ distribution of K giant stars excluded all samples inside the thick blue polygon defined in the top-right panel. \emph{Bottom-right panel}: The \teff\ vs. \logg\ distribution of the stars inside the thick blue polygon, in which most of the red clump stars are located.}\label{fig:tefflogg}
\end{minipage}
\end{center}
\end{figure}

The top-left panel of figure~\ref{fig:tefflogg} shows the location of the red clump stars in \teff\ vs. \logg\ plane. We use a polygon (red lines) to select the possible red clump stars from their \teff\ and \logg. However, for the majority of the identified K giant stars, there is no measurement of \teff\ and \logg. We turn to use alternative quantities directly measured from the spectra. The top-right panel shows the $EW_{Mg_b}$ vs. \fehlmd\ for the selected K giant stars (black filled contours). The possible red clump stars inside the red polygon are mostly concentrated into a narrower region (yellow contours). For simplicity, we use the thick blue polygon shown in the top-right panel to locate the red clump stars. Subsequently, we remove all the stars that fall in the polygon to exclude the red clump star contamination. After the exclusion of the possible red clump stars, the \teff\ vs. \logg\ distribution of the rest of the K giant stars is shown in the bottom-left panel. The clump centered at \teff$\sim4900$ and \logg$\sim2.5$ in the top-left panel is now almost disappeared. On the other hand, the excluded stars located within the thick blue polygon are concentrated as a clump in the \teff\ vs. \logg\ plane, as shown in the bottom-right panel.

A quantitative assessment of the performance of the red clump star exclusion method is very difficult since we cannot definitely determine whether an individual field star is a red clump star. However, figure~\ref{fig:tefflogg} shows that, qualitatively, the simple polygon exclusion works quite well in the removing of the red clump stars. This is sufficient for the spatial overview of the K giant stars, losing a small fraction of K giant stars being mistakenly classified as red clump stars and keeping a smaller fraction of contaminated red clump stars in the sample. The distances for the contaminated red clump stars are likely underestimated because that the RGB/SGB stars with same $J-K$ and \feh\ have fainter absolute magnitude than that of the red clump stars. An elegant classification of the red clump stars deserves an another specific project and beyond the scope of this work.

\end{document}